\documentclass[twocolumn,tighten]{aastex63}
\usepackage{graphicx}
\graphicspath{{./figures/}{./}}

\usepackage{listings}
\usepackage{enumitem}
\usepackage{lineno}
\usepackage{soul}
\usepackage{comment}
\input{hyperlink-year-only-natbib-patch}

\usepackage{soul} 
\usepackage{amsmath}
\usepackage{amssymb}
\usepackage{xspace}
\usepackage{xifthen}

\newcommand{\ie}{i.e.\xspace}
\newcommand{\eg}{e.g.\xspace}
\newcommand{\etc}{etc.\xspace}

\newcommand{\RESPONSE}[1]{{}}
\newcommand{\REMOVED}[1]{{}}

\newcommand{\name}{NGC~55-dw1\xspace}
\newcommand{\gold}{Y6 Gold\xspace}

\mathchardef\mhyphen="2D

\newcommand{\roughly}{\ensuremath{ {\sim}\,} }

\newlength{\dhatheight}

\newcommand{\code}[1]{\texttt{#1}\xspace}

\newcommand{\unit}[1]{\ensuremath{\mathrm{\,#1}}\xspace}
\newcommand{\Gyr}{\unit{Gyr}}

\newcommand{\pc}{\unit{pc}}
\newcommand{\kpc}{\unit{kpc}}
\newcommand{\Mpc}{\unit{Mpc}}
\newcommand{\second}{\unit{s}}

\newcommand{\Msolar}{\unit{M_\odot}}
\newcommand{\Msun}{\unit{M_\odot}}

\newcommand{\Lsolar}{\unit{L_\odot}}

\newcommand{\magn}{\unit{mag}}
\newcommand{\mmag}{\unit{mmag}}
\newcommand{\e}{\unit{e^{-}}}

\newcommand{\secref}[1]{Section~\ref{sec:#1}}
\newcommand{\appref}[1]{Appendix~\ref{app:#1}}
\newcommand{\tabref}[1]{Table~\ref{tab:#1}}
\newcommand{\figref}[1]{Figure~\ref{fig:#1}}

\newcommand{\bandvar}[2][]{%
  \ifthenelse{\isempty{#1}}{\var{#2}}{\var{#2\_#1}}%
}

\newcommand{\LCDM}{\ensuremath{\rm \Lambda CDM}\xspace}
\newcommand{\modulus}{\ensuremath{m - M}\xspace}
\newcommand{\mM}{\modulus}
\newcommand{\ra}{{\ensuremath{\alpha_{2000}}}\xspace}
\newcommand{\dec}{{\ensuremath{\delta_{2000}}}\xspace}
\newcommand{\age}{{\ensuremath{\tau}}\xspace}
\newcommand{\metal}{{\ensuremath{Z}}\xspace}
\newcommand{\major}{\ensuremath{a_h}\xspace}
\newcommand{\feh}{{\ensuremath{\rm [Fe/H]}}\xspace}
\newcommand{\ellip}{\ensuremath{\epsilon}\xspace}
\newcommand{\PA}{\ensuremath{\mathrm{P.A.}}\xspace}

\newcommand{\SExtractor}{\code{Source{\allowbreak}Extractor}}

\newcommand{\HEALPix}{\code{HEALPix}}
\newcommand{\healpix}{\HEALPix}
\newcommand{\nside}{\code{nside}}

\newcommand{\emcee}{\code{emcee}}
\newcommand{\ugali}{\code{ugali}}
\newcommand{\simple}{\code{simple}}
\newcommand{\var}[1]{\ensuremath{\texttt{\MakeUppercase{#1}}}\xspace}

\newcommand{\rhalf}{\ensuremath{ r_{\rm h} }\xspace}

\providecommand\physrep{\ref@jnl{Phys.~Rep.}}%
\providecommand\apjs{\ref@jnl{ApJS}}%
\providecommand{\jcap}{\ref@jnl{JCAP}}%

\defcitealias{DR1:2018}{DES Collaboration (2018)}

\makeatletter
\providecommand*{\input@path}{}
\g@addto@macro\input@path{{./tables/}}
\makeatother

\usepackage{makecell}

\usepackage{hyperref}

\shorttitle{Distant Dwarfs in DES Y6}
\shortauthors{McNanna et al.}

\begin{document}

\reportnum{DES-2023-772}
\reportnum{FERMILAB-PUB-23-478-PPD}

\title{A search for faint resolved galaxies beyond the Milky Way in DES Year 6: A new faint, diffuse dwarf satellite of NGC~55}

\author[0000-0001-5435-7820]{M.~McNanna}
\affiliation{Department of Physics, University of Wisconsin-Madison, 1150 University Avenue, Madison, WI 53706, USA}

\author[0000-0001-8156-0429]{K.~Bechtol}
\affiliation{Department of Physics, University of Wisconsin-Madison, 1150 University Avenue, Madison, WI 53706, USA}

\author[0000-0003-3519-4004]{S.~Mau}
\affiliation{Department of Physics, Stanford University, 382 Via Pueblo Mall, Stanford, CA 94305, USA}
\affiliation{Kavli Institute for Particle Astrophysics \& Cosmology, P. O. Box 2450, Stanford University, Stanford, CA 94305, USA}

\author[0000-0002-1182-3825]{E.~O.~Nadler}
\affiliation{Carnegie Observatories, 813 Santa Barbara Street, Pasadena, CA 91101, USA}
\affiliation{Department of Physics $\&$ Astronomy, University of Southern California, Los Angeles, CA, 90007, USA}

\author[0000-0003-1162-7346]{J.~Medoff}
\affiliation{Department of Astronomy and Astrophysics, University of Chicago, Chicago, IL 60637, USA}

\author[0000-0001-8251-933X]{A.~Drlica-Wagner}
\affiliation{Department of Astronomy and Astrophysics, University of Chicago, Chicago, IL 60637, USA}
\affiliation{Fermi National Accelerator Laboratory, P. O. Box 500, Batavia, IL 60510, USA}
\affiliation{Kavli Institute for Cosmological Physics, University of Chicago, Chicago, IL 60637, USA}

\author[0000-0003-1697-7062]{W.~Cerny}
\affiliation{Department of Astronomy, Yale University, New Haven, CT 06520, USA}

\author[0000-0002-1763-4128]{D.~Crnojevi\'{c}}
\affiliation{Department of Physics and Astronomy, University of Tampa, 401 West Kennedy Boulevard, Tampa, FL 33606, USA}

\author[0000-0001-9649-4815]{B.~Mutlu-Pakd{\i}l}
\affiliation{Department of Physics and Astronomy, Dartmouth College, Hanover, NH 03755, USA}

\author[0000-0003-4341-6172]{A.~K.~Vivas}
\affiliation{Cerro Tololo Inter-American Observatory/NSF's NOIRLab, Casilla 603, La Serena, Chile}

\author[0000-0002-6021-8760]{A.~B.~Pace}
\affiliation{McWilliams Center for Cosmology, Carnegie Mellon University, 5000 Forbes Ave, Pittsburgh, PA 15213, USA}

\author[0000-0002-3936-9628]{J.~L.~Carlin}
\affiliation{Vera C. Rubin Observatory/AURA, 950 N Cherry Ave, Tucson, AZ 85719 USA}

\author[0000-0002-1693-3265]{M.~L.~M.~Collins}
\affiliation{Department of Physics, University of Surrey, Guildford, GU2 7XH, UK}

\author[0000-0001-6957-1627]{P.~S.~Ferguson}
\affiliation{Department of Physics, University of Wisconsin-Madison, 1150 University Avenue, Madison, WI 53706, USA}

\author[0000-0003-3835-2231]{D.~Mart\'inez-Delgado} 
\affiliation{Instituto de Astrof\'isica de Andaluc\'ia, CSIC, Glorieta de la Astronom\'ia, E-18080, Granada, Spain}

\author[0000-0002-9144-7726]{C.~E.~Mart\'inez-V\'azquez}
\affiliation{Gemini Observatory, NSF’s NOIRLab, 670 N. A’ohoku Place, Hilo, HI 96720, USA}

\author[0000-0002-8282-469X]{N.~E.~D.~Noel}
\affiliation{Department of Physics, University of Surrey, Guildford, GU2 7XH, UK}

\author[0000-0001-5805-5766]{A.~H.~Riley}
\affiliation{Institute for Computational Cosmology, Department of Physics, Durham University, South Road, Durham DH1 3LE, UK}

\author[0000-0003-4102-380X]{D.~J.~Sand}
\affil{Steward Observatory, University of Arizona, 933 North Cherry Avenue, Tucson, AZ 85721-0065, USA}

\author[0000-0003-2599-7524]{A.~Smercina}
\affiliation{Department of Astronomy, University of Washington, Box 351580, U.W., Seattle, WA 98195-1580, USA}

\author[0000-0002-9599-310X]{E.~Tollerud}
\affiliation{Space Telescope Science Institute, 3700 San Martin Dr., Baltimore, MD 21218, USA}

\author[0000-0003-2229-011X]{R.~H.~Wechsler}
\affiliation{Department of Physics, Stanford University, 382 Via Pueblo Mall, Stanford, CA 94305, USA}
\affiliation{Kavli Institute for Particle Astrophysics \& Cosmology, P. O. Box 2450, Stanford University, Stanford, CA 94305, USA}
\affiliation{SLAC National Accelerator Laboratory, Menlo Park, CA 94025, USA}

\author{T.~M.~C.~Abbott}
\affiliation{Cerro Tololo Inter-American Observatory/NSF's NOIRLab, Casilla 603, La Serena, Chile}

\author{M.~Aguena}
\affiliation{Laborat\'orio Interinstitucional de e-Astronomia - LIneA, Rua Gal. Jos\'e Cristino 77, Rio de Janeiro, RJ - 20921-400, Brazil}

\author{O.~Alves}
\affiliation{Department of Physics, University of Michigan, Ann Arbor, MI 48109, USA}

\author{D.~Bacon}
\affiliation{Institute of Cosmology and Gravitation, University of Portsmouth, Portsmouth, PO1 3FX, UK}

\author[0000-0003-4383-2969]{C.~R.~Bom}
\affiliation{Centro Brasileiro de Pesquisas F\'isicas, Rua Dr. Xavier Sigaud 150, 22290-180 Rio de Janeiro, RJ, Brazil}

\author[0000-0002-8458-5047]{D.~Brooks}
\affiliation{Department of Physics \& Astronomy, University College London, Gower Street, London, WC1E 6BT, UK}

\author{D.~L.~Burke}
\affiliation{Kavli Institute for Particle Astrophysics \& Cosmology, P. O. Box 2450, Stanford University, Stanford, CA 94305, USA}
\affiliation{SLAC National Accelerator Laboratory, Menlo Park, CA 94025, USA}

\author[0000-0002-3690-105X]{J.~A.~Carballo-Bello}
\affiliation{Instituto de Alta Investigaci\'on, Sede Esmeralda, Universidad de Tarapac\'a, Av. Luis Emilio Recabarren 2477, Iquique, Chile}

\author[0000-0003-3044-5150]{A.~Carnero~Rosell}
\affiliation{Instituto de Astrofisica de Canarias, E-38205 La Laguna, Tenerife, Spain}
\affiliation{Laborat\'orio Interinstitucional de e-Astronomia - LIneA, Rua Gal. Jos\'e Cristino 77, Rio de Janeiro, RJ - 20921-400, Brazil}
\affiliation{Universidad de La Laguna, Dpto. Astrof\'isica, E-38206 La Laguna, Tenerife, Spain}

\author[0000-0002-3130-0204]{J.~Carretero}
\affiliation{Institut de F\'isica d'Altes Energies (IFAE), The Barcelona Institute of Science and Technology, Campus UAB, 08193 Bellaterra (Barcelona) Spain}

\author{L.~N.~da Costa}
\affiliation{Laborat\'orio Interinstitucional de e-Astronomia - LIneA, Rua Gal. Jos\'e Cristino 77, Rio de Janeiro, RJ - 20921-400, Brazil}

\author[0000-0002-4213-8783]{T.~M.~Davis}
\affiliation{School of Mathematics and Physics, University of Queensland,  Brisbane, QLD 4072, Australia}

\author[0000-0001-8318-6813]{J.~De~Vicente}
\affiliation{Centro de Investigaciones Energ\'eticas, Medioambientales y Tecnol\'ogicas (CIEMAT), Madrid, Spain}

\author[0000-0002-8357-7467]{H.~T.~Diehl}
\affiliation{Fermi National Accelerator Laboratory, P. O. Box 500, Batavia, IL 60510, USA}

\author{P.~Doel}
\affiliation{Department of Physics \& Astronomy, University College London, Gower Street, London, WC1E 6BT, UK}

\author{I.~Ferrero}
\affiliation{Institute of Theoretical Astrophysics, University of Oslo. P.O. Box 1029 Blindern, NO-0315 Oslo, Norway}

\author[0000-0003-4079-3263]{J.~Frieman}
\affiliation{Fermi National Accelerator Laboratory, P. O. Box 500, Batavia, IL 60510, USA}
\affiliation{Kavli Institute for Cosmological Physics, University of Chicago, Chicago, IL 60637, USA}

\author[0000-0002-3730-1750]{G.~Giannini}
\affiliation{Institut de F\'isica d'Altes Energies (IFAE), The Barcelona Institute of Science and Technology, Campus UAB, 08193 Bellaterra (Barcelona) Spain}

\author[0000-0003-3270-7644]{D.~Gruen}
\affiliation{University Observatory, Faculty of Physics, Ludwig-Maximilians-Universit\"at, Scheinerstr. 1, 81679 Munich, Germany}

\author[0000-0003-0825-0517]{G.~Gutierrez}
\affiliation{Fermi National Accelerator Laboratory, P. O. Box 500, Batavia, IL 60510, USA}

\author{R.~A.~Gruendl}
\affiliation{Center for Astrophysical Surveys, National Center for Supercomputing Applications, 1205 West Clark St., Urbana, IL 61801, USA}
\affiliation{Department of Astronomy, University of Illinois at Urbana-Champaign, 1002 W. Green Street, Urbana, IL 61801, USA}

\author{S.~R.~Hinton}
\affiliation{School of Mathematics and Physics, University of Queensland,  Brisbane, QLD 4072, Australia}

\author{D.~L.~Hollowood}
\affiliation{Santa Cruz Institute for Particle Physics, Santa Cruz, CA 95064, USA}

\author[0000-0002-6550-2023]{K.~Honscheid}
\affiliation{Center for Cosmology and Astro-Particle Physics, The Ohio State University, Columbus, OH 43210, USA}
\affiliation{Department of Physics, The Ohio State University, Columbus, OH 43210, USA}

\author[0000-0001-5160-4486]{D.~J.~James}
\affiliation{ASTRAVEO LLC, PO Box 1668, MA 01931}  
\affiliation{Applied Materials, Inc., 35 Dory Road, Gloucester, MA 01930}

\author[0000-0003-0120-0808]{K.~Kuehn}
\affiliation{Australian Astronomical Optics, Macquarie University, North Ryde, NSW 2113, Australia}
\affiliation{Lowell Observatory, 1400 Mars Hill Rd, Flagstaff, AZ 86001, USA}

\author[0000-0003-0710-9474]{J.~L.~Marshall}
\affiliation{George P. and Cynthia Woods Mitchell Institute for Fundamental Physics and Astronomy, and Department of Physics and Astronomy, Texas A\&M University, College Station, TX 77843,  USA}

\author[0000-0001-9497-7266]{J.~Mena-Fern{\'a}ndez}
\affiliation{Centro de Investigaciones Energ\'eticas, Medioambientales y Tecnol\'ogicas (CIEMAT), Madrid, Spain}

\author[0000-0002-6610-4836]{R.~Miquel}
\affiliation{Instituci\'o Catalana de Recerca i Estudis Avan\c{c}ats, E-08010 Barcelona, Spain}
\affiliation{Institut de F\'isica d'Altes Energies (IFAE), The Barcelona Institute of Science and Technology, Campus UAB, 08193 Bellaterra (Barcelona) Spain}

\author{M.~E.~S.~Pereira}
\affiliation{Hamburger Sternwarte, Universit\"{a}t Hamburg, Gojenbergsweg 112, 21029 Hamburg, Germany}

\author[0000-0001-9186-6042]{A.~Pieres}
\affiliation{Laborat\'orio Interinstitucional de e-Astronomia - LIneA, Rua Gal. Jos\'e Cristino 77, Rio de Janeiro, RJ - 20921-400, Brazil}
\affiliation{Observat\'orio Nacional, Rua Gal. Jos\'e Cristino 77, Rio de Janeiro, RJ - 20921-400, Brazil}

\author[0000-0002-2598-0514]{A.~A.~Plazas~Malag\'on}
\affiliation{Kavli Institute for Particle Astrophysics \& Cosmology, P. O. Box 2450, Stanford University, Stanford, CA 94305, USA}
\affiliation{SLAC National Accelerator Laboratory, Menlo Park, CA 94025, USA}

\author[0000-0002-1594-1466]{J.~D.~Sakowska}
\affiliation{Department of Physics, University of Surrey, Guildford, GU2 7XH, UK}

\author[0000-0002-9646-8198]{E.~Sanchez}
\affiliation{Centro de Investigaciones Energ\'eticas, Medioambientales y Tecnol\'ogicas (CIEMAT), Madrid, Spain}

\author[0000-0003-3054-7907]{D.~Sanchez Cid}
\affiliation{Centro de Investigaciones Energ\'eticas, Medioambientales y Tecnol\'ogicas (CIEMAT), Madrid, Spain}

\author{B.~Santiago}
\affiliation{Instituto de F\'isica, UFRGS, Caixa Postal 15051, Porto Alegre, RS - 91501-970, Brazil}
\affiliation{Laborat\'orio Interinstitucional de e-Astronomia - LIneA, Rua Gal. Jos\'e Cristino 77, Rio de Janeiro, RJ - 20921-400, Brazil}

\author[0000-0002-1831-1953]{I.~Sevilla-Noarbe}
\affiliation{Centro de Investigaciones Energ\'eticas, Medioambientales y Tecnol\'ogicas (CIEMAT), Madrid, Spain}

\author[0000-0002-3321-1432]{M.~Smith}
\affiliation{School of Physics and Astronomy, University of Southampton,  Southampton, SO17 1BJ, UK}

\author[0000-0003-1479-3059]{G.~S.~Stringfellow}
\affiliation{Center for Astrophysics and Space Astronomy, University of Colorado, 389 UCB, Boulder, CO 80309-0389, USA}

\author[0000-0002-7047-9358]{E.~Suchyta}
\affiliation{Computer Science and Mathematics Division, Oak Ridge National Laboratory, Oak Ridge, TN 37831}

\author{M.~E.~C.~Swanson}
\affiliation{National Center for Supercomputing Applications, 1205 West Clark St., Urbana, IL 61801, USA}

\author[0000-0003-1704-0781]{G.~Tarle}
\affiliation{Department of Physics, University of Michigan, Ann Arbor, MI 48109, USA}

\author{N.~Weaverdyck}
\affiliation{Department of Physics, University of Michigan, Ann Arbor, MI 48109, USA}
\affiliation{Lawrence Berkeley National Laboratory, 1 Cyclotron Road, Berkeley, CA 94720, USA}

\author{P.~Wiseman}
\affiliation{School of Physics and Astronomy, University of Southampton,  Southampton, SO17 1BJ, UK}

\collaboration{66}{(DES \& DELVE Collaboration)}

\correspondingauthor{Mitch McNanna}
\email{mcnanna@wisc.edu}

\begin{abstract}
We report results from a systematic wide-area search for faint dwarf galaxies at heliocentric distances from 0.3 to 2\Mpc using the full six years of data from the Dark Energy Survey (DES). 
Unlike previous searches over the DES data, this search specifically targeted a field population of faint galaxies located beyond the Milky Way virial radius.
We derive our detection efficiency for faint, resolved dwarf galaxies in the Local Volume with a set of synthetic galaxies and expect our search to be complete to $M_V \roughly (-7, -10)\magn$ for galaxies at $D = (0.3, 2.0)\Mpc$ respectively. 
We find no new field dwarfs in the DES footprint, but we report the discovery of one high-significance candidate dwarf galaxy at a distance of $2.2\substack{+0.05\\-0.12}\Mpc$, a potential satellite of the Local Volume galaxy NGC~55, separated by $47$~arcmin (physical separation as small as 30\kpc).
We estimate this dwarf galaxy to have an absolute V-band magnitude of $-8.0\substack{+0.5\\-0.3}\magn$ and an azimuthally averaged physical half-light radius of $2.2\substack{+0.5\\-0.4}\kpc$, making this one of the lowest surface brightness galaxies ever found with $\mu = 32.3\magn\,{\rm arcsec}^{-2}$. 
This is the largest, most diffuse galaxy known at this luminosity, suggesting possible tidal interactions with its host. 

\end{abstract}

\section{Introduction}

Dwarf galaxies are the most abundant galaxies in the Universe, and their demographics offer a unique probe into galaxy formation and feedback processes, reionization, and the nature of dark matter.
The brightest Local Group (LG) galaxies were historically discovered predominantly in visual searches of photographic plates \citep{Shapley:1938a, Shapley:1938b, Harrington:1950, Wilson:1955, Cannon:1977, Irwin:1990, Ibata:1994}.
Large digital sky surveys have since allowed for fainter systems to be discovered using statistical matched-filter techniques, identifying faint dwarf galaxies as arcminute-scale overdensities of old, metal poor stars \citep{Willman:2005a, Willman:2005b, Zucker:2006a, Zucker:2006b, Belokurov:2006, Belokurov:2007, Belokurov:2008, Belokurov:2009, Belokurov:2010, Grillmair:2006, Grillmair:2009, Sakamoto:2006, Irwin:2007, Walsh:2007}.
Searches using these matched-filter techniques have been applied to the current generation of wide imaging surveys to detect yet fainter and more distant systems \citep{Bechtol:2015, Drlica-Wagner:2015, Koposov:2015, Koposov:2018, Kim:2015d, Kim:2015b, Kim:2015c, martin_2015_hydra_ii, Laevens:2015a, Laevens:2015b, Torrealba:2016a, Torrealba:2016b, Torrealba:2018a, Torrealba:2019, Homma:2016, Homma:2017, Homma:2019,  Luque:2017, Mau:2020, Cerny:2021b, Cerny:2022, Cerny:2023,Smith:2023}.

Ultra-faint dwarf galaxies ($M_V \gtrsim -7.7$, \citealt{Simon:2019}) are the most dark matter-dominated systems known and represent the extreme limit of the galaxy formation process, likely inhabiting the lowest-mass dark matter halos capable of hosting star formation \citep{Jethwa:2018,Wheeler:2019,Nadler:2020,Applebaum:2021}.
Recent systematic searches for ultra-faint Milky Way (MW) satellite galaxies over $\roughly 80\%$ of the sky have allowed for robust inferences about the population of such galaxies within the virial radius of the MW \citep{Koposov:2008, Drlica-Wagner:2020}.
This census has allowed for the first constraints on the galaxy-halo connection for dark matter halos below $10^8 \: M_\odot$, including evidence for the statistical impact of the Large Magellanic Cloud (LMC) on the MW satellite population \citep{Nadler:2020}, and limits on the properties of several alternative dark matter models \citep{Newton:2018, Newton:2021, Kim:2018, Nadler:2021, Mau:2022}.  

However, the population of LG galaxies beyond the MW virial radius (300\kpc) is less explored.
Dwarf galaxies dominate the universe by number, yet a precise census of these objects remains challenging due to their inherently faint nature and the limited sensitivity of observational surveys.
In the nearby universe, these low-luminosity dwarf galaxies are detected in optical imaging surveys as arcminute-scale statistical overdensities of individually resolved stars. 
Previous searches for distant dwarf galaxies have primarily been targeted searches of the halos of larger host galaxies, typically out to their virial radii. 
A satellite census has been performed for M31 \citep{McConnachie:2008, McConnachie:2009, Martin:2009, Martin:2013, Martin:2016} and for several other large galaxies within the Local Volume \citep[$D \lesssim 11 \Mpc$;][]{Taylor:2018, Crnojevic:2016, Muller:2019, Chiboucas:2013, Smercina:2018, Merritt:2014, Bennet:2019, Bennet:2020, Sand:2014, Toloba:2016, Martinez-Delgado:2021, Mutlu-Pakdil:2022, Carlsten:2022}.
Targeted searches have also been performed around smaller, Magellanic Cloud analogue galaxies more nearby in the Local Group as part of the MADCASH (Magellanic Analog Dwarf Companions and Stellar Halos) project \citep{Carlin:2016, Carlin:2021} and the DEEP component of the DECam Local Volume Exploration (DELVE-DEEP, \citealt{Drlica-Wagner:2021}).

Recent studies of the LMC and its impact on the MW satellite population have indicated that these targeted searches are likely to be fruitful, since the LMC fell into the MW with its own satellite population \citep{Kallivayalil:2018, Patel:2020, Nadler:2020}. However, in addition to the prediction of faint satellites around field galaxies in the Local Volume \citep{Dooley:2017a}, cosmological zoom-in simulations of MW- and LG-like systems also predict the existence of low-mass halos outside the virial radii of a larger host \citep{Garrison-Kimmel:2014, Garrison-Kimmel:2019a, Nadler:2020, Joshi:2023}.
These isolated halos are either ``field" halos, having never passed within the virial radius of a larger host halo, ``splashback" halos which have orbited once within the virial radius of a larger host but today reside outside \citep{Adhikari:2014, Diemer:2014, More:2015}, or ``Hermean" halos that passed through the halos of both the Milky Way and M31 at early times \citep{Newton:2022}.
Known highly isolated dwarf galaxies around the LG are catalogued in \citet{Martinez-Delgado:2018}, and are all relatively bright ($M_V \lesssim -10\magn$). 
 
Simulated field dwarf populations agree with observations at the bright end, matching subhalo mass functions for $M_* \geq 10^6 \Msun$ in hydrodynamic simulations including baryonic feedback \citep{Fattahi:2016,Sawala:2016,Garrison-Kimmel:2019a,Applebaum:2021}; however, an unobserved population of low-mass dwarf galaxies with $M_* = 10^5 - 10^6\Msun$ also exists in these simulations \citep{Garrison-Kimmel:2019a, Fattahi:2020}.
Discovering and characterizing these low-mass isolated field dwarfs could lend insight into the ``too-big-to-fail" problem (TBTF, \citealt{Boylan-Kolchin:2011, Boylan-Kolchin:2012}), which is present for field galaxies beyond the MW and beyond the LG entirely \citep{Papastergis:2015, Papastergis:2016}. 
The existence of low-density systems in question for TBTF tensions could be interpreted as evidence for alternative dark matter models; in particular, strongly self-interacting dark matter models (SIDM) diversify low-mass halo populations relative to CDM, predicting both underdense and overdense outliers among LG isolated dwarfs \citep{Robles:2017, Fitts:2019, Daneng:2022}. 
The Tucana dwarf galaxy, which may be a tidally affected splashback system of M31 \citep{Santos-Santos:2023}, was originally thought to be an example of such an overly dense system \citep{Gregory:2019, Fraternali:2009}, although more recent analyses have found a central density profile consistent with other LG dwarfs \citep{Taibi:2020}. 

Blind searches for resolved dwarf galaxies in the LG field ($D \approx 0.3 - 2 \Mpc$) offer separate challenges from targeted satellite searches. 
In particular, the 3D search volume is much larger. 
Lacking a strong prior on the heliocentric distance distribution of these isolated systems requires us to scan over the line-of-sight distance at each sky location. 
Additionally, a field search requires much broader sky coverage.
Until recently, wide optical surveys have not been sensitive enough to detect faint dwarf galaxies at these distances. 
However, recent discoveries of relatively isolated LG ultra-faint dwarf galaxies \citep{Martinez-Delgado:2022, Collins:2022, Collins:2023, McQuinn:2023, McQuinn:2023b} have been made in imaging from the DESI Legacy Imaging Survey \citep{Dey:2019}, a DECam survey reaching depths of $\roughly 24\magn$ in the optical bands, comparable to the apparent magnitude of red giant branch (RGB) stars in dwarf galaxies $\gtrsim 1\Mpc$. 
The recently processed data from the six-year Dark Energy Survey (DES Y6), another DECam survey $\roughly 0.5\magn$ deeper than the DESI Legacy Imaging Survey \citep{Abbott:2021}, will similarly be sensitive to the brightest members of this distant population. 

Here we report on a wide-area search for a field population of faint dwarf galaxies beyond the MW virial radius. 
We performed a search over the entire 5000~deg$^2$ DES footprint using the same search algorithm employed in previous wide-area DES searches \citep{Drlica-Wagner:2020} as well as in the recent discoveries of several new nearby ultra-faint stellar systems \citep[e.g.,][]{Mau:2020, Cerny:2021a, Cerny:2021b, Cerny:2023} (\secref{wide}).
We optimized the algorithm for the detection of more distant systems by including multi-band photometry and searching the footprint on a finer spatial grid, and injected synthetic galaxies into the DES data to quantify the search sensitivity.
The search yielded a single high-significance candidate, designated DES~J0015-3825, based on a stellar population consistent with the tip of the red giant branch of an old, metal-poor stellar population at a distance of $\roughly 2\Mpc$ (\secref{cand}).
We use deeper follow-up DECam images of the candidate to confirm and characterize it. 
The proximity of DES~J0015-3825 to the LMC-mass galaxy NGC~55 suggests the presence of a low luminosity central-satellite system and possible tidal interactions between the two galaxies; we therefore refer to the candidate dwarf galaxy as \name throughout this paper. 
Finally, we discuss the implications for the total galaxy population within 2\Mpc and the outlook for searches with future wide-area imaging surveys (\secref{discussion}).

\section{Wide-Area Search}
\label{sec:wide}

\subsection{Data}
\label{sec:wide_data}

The wide-area search used data from the Dark Energy Survey (DES), an optical and near-infrared ground-based wide-area imaging survey covering $\roughly 5000 \deg^2$ of the southern high-galactic latitude sky. 
DES utilizes the Dark Energy Camera (DECam; \citealt{Honscheid:2008}; \citealt{Flaugher:2015}), a $3 \deg^2$ field-of-view camera installed at the prime focus of the 4 m Blanco telescope at the Cerro Tololo Inter-American Observatory. 
In this paper we made use of the full six years of DES wide-area survey observations (DES Y6). 
This dataset was released publicly as the second public data release (DR2) of DES data \citep{Abbott:2021}. 
Image reduction and processing were performed by the DES Data Management system (DESDM, \citealt{Morganson:2018}) at the National Center for Supercomputing Applications (NCSA).  
The processed images from this pipeline are used to build a coadded catalog of astronomical objects, with PSF model fitting performed by \code{PSFEx} \citep{Bertin:2011} and source detection and measurement performed by \code{SourceExtractor} \citep{Bertin:1996}. 
Internal photometric calibration was performed using \code{FGCM} \citep{Burke:2018} to obtain a uniformity of better than 2 \mmag across the survey footprint \citep{Rykoff:2023}. 
The estimated coadded imaging depth in the optical bands, defined as S/N$=10$, is 24.7, 24.4, and 23.8\magn in each of $gri$ respectively \citep{Abbott:2021}. 

The stellar sample used in this work comes from a preliminary version of a \gold dataset that expands upon data products included in DES DR2, analogous to the ensemble of Y3 Gold data products for the first three years of DES data \citep{DESY3Gold}.
Photometric measurements use the multi-epoch, multi-band fitting algorithm \code{fitvd} based on a combined fit of $grizY$ measurements, considering the PSF model results for the stellar sample.
Objects for the stellar sample are selected using a morphological classification based on a comparison between PSF and extended models with \code{fitvd}, using a threshold in the size parameter \code{BDF\_T} that maximizes the Matthews Correlation Coefficient in each signal-to-noise bin.
For the $0.2\%$ subset of objects without robust \code{fitvd} extended model measurements, we use the \SExtractor \code{WAVG\_SPREAD\_MODEL\_I} or \code{SPREAD\_MODEL\_I} morphological measurements for star-galaxy classification as described in \citet{Abbott:2021}.
We select quality objects using the \code{FLAGS\_GOLD} = 0 criteria.
The effects of interstellar extinction are taken into account by de-reddening flux measurements using the SFD dust map \citep{Schlegel:1998} as described in \citet{Abbott:2021}.

We limit the analysis to regions of the DES footprint that have at least 50\% coverage in the intersection of the $griz$ bands, and at least 2 exposures in each of the $griz$ bands, as measured within arcminute-scale spatial pixels; survey properties are represented using \healpix maps at resolution \nside = 4096 \citep{Abbott:2021}.
The total area of the footprint used in the search is 4838 square degrees in a single contiguous region.

\subsection{Search Algorithm}
\label{sec:algorithm}

Our search employed an automated algorithm similar to previous works searching for ultra-faint galaxies \citep[\eg,][]{Drlica-Wagner:2020, Mau:2020, Cerny:2022}, but modified to be more sensitive to distant dwarf galaxies. 
This search algorithm, \simple\footnote{\url{https://github.com/sidneymau/simple/tree/simple3}}, uses a simple isochrone filter to remove the foreground field of Milky Way stars and enhance the contrast of substructures at a given distance. \simple is inspired by the matched-filter methods of \citet{Koposov:2008} and \citet{Walsh:2009}, and this specific implementation builds on the technique described by \citet{Bechtol:2015} and \citet{Drlica-Wagner:2015}.  

Since the member stars of the distant systems we aim to discover sit near the depth limit of our data, we used multi-band photometry to help facilitate star-galaxy classification beyond the morphological classifiers included in the DES Y6 data. 
In particular, we require that objects be detected in all of the $gri$ bands. 
This is in contrast to previous implementations of the \simple algorithm which only used photometry from two bands, generally $g$ and $r$ (\eg, \citealt{Drlica-Wagner:2020}, \citealt{Cerny:2022}).
As a basic quality selection, we required that objects be brighter than $r = 24.5\magn$ and $i = 24.25\magn$, consistent with previous implementations of \simple on DES data and adjusted to reflect the increased depth of DES Y6. 
We further required that objects lie near the stellar locus in color-color space.
Specifically, we required the straight-line distance $d$ between an object's ($g-r$, $r-i$) coordinate in color-color space and the best-fit line to the stellar locus satisfy $d < \sqrt{0.15^2 + \sigma_g^2 + \sigma_r^2 + \sigma_i^2}$, where $\sigma_g$ is the uncertainty in the $g$-band magnitude and so on for the other bands. 
The selection was chosen to be inclusive enough to ensure almost all correctly classified stars will still be included while simultaneously excluding point-like sources whose color clearly identifies them as galaxies. 

A matched-filter search for old, metal-poor stars was performed over distance moduli ranging from $22.5 \leq \modulus \leq 26.5\magn$ in steps of 0.5\magn, corresponding to a search between heliocentric distances of $316 \leq D \leq 2000\kpc$. 
This range begins at the MW virial radius ($\sim 300\kpc$) and ends at the approximate distance modulus beyond which too few RGB stars will lie above the survey's depth limit to enable confident detection. 
At each distance modulus, we selected stars with $g$-, $r$-, and $i$-band magnitudes consistent with synthetic isochrones of \citet{Bressan:2012} with metallicity $\metal = 0.0001$ and age $\age = 12\Gyr$. 
In particular, we applied two separate isochrone filters: one using $g-r$ color vs $g$-band magnitude, and another using $r-i$ color vs $r$-band magnitude. 
In each case we required that the color difference between each star and the template isochrone be within 0.1\magn accounting for uncertainties in the magnitudes, \ie $\Delta (g-r) < \sqrt{0.1^2 + \sigma_g^2 + \sigma_r^2}$.

The DES footprint was partitioned into $\nside=32$ \healpix pixels ($\roughly 3.4 \deg^2$), and each pixel was analyzed individually. 
For each pixel and distance modulus step, the color and isochrone filters were applied as previously described to the central \healpix pixel along with the eight surrounding pixels, creating a map of the filtered stellar density field in the region of interest. 
The eight surrounding pixels are necessary to more accurately estimate the average stellar density in the region. 
This density field was smoothed by a Gaussian kernel with $\sigma = 1\arcmin$.
Crucially, the spatial bin size for the density map and the smoothing kernel were smaller in our implementation of \simple relative to previous searches to aid in the detection of distant galaxies whose angular size is small relative to more nearby MW satellites with similar structural parameters. 
We perform a first pass search for local density peaks in the smoothed map by iteratively raising a density threshold until there are fewer than 10 disconnected peaks above the threshold. 
For each peak identified, we estimate the local field density using an annulus between $18\arcmin$ and $30\arcmin$ centered on the peak.
We account here for the survey coverage which is mapped at arcmin$^2$ scales. 
We then iterate through circular apertures with radii from $0.18\arcmin$ to $18\arcmin$ and compute the Poisson significance for the observed stellar count within the aperture relative to the local field density, identifying the angular size of the aperture which maximizes significance. 
Spatially coincident peaks at different distance moduli are consolidated, identifying the modulus with the largest significance as an estimate of distance.

\subsection{Results}
\label{wide_results}

\begin{figure*}
\centering
\includegraphics[width=1.\textwidth]{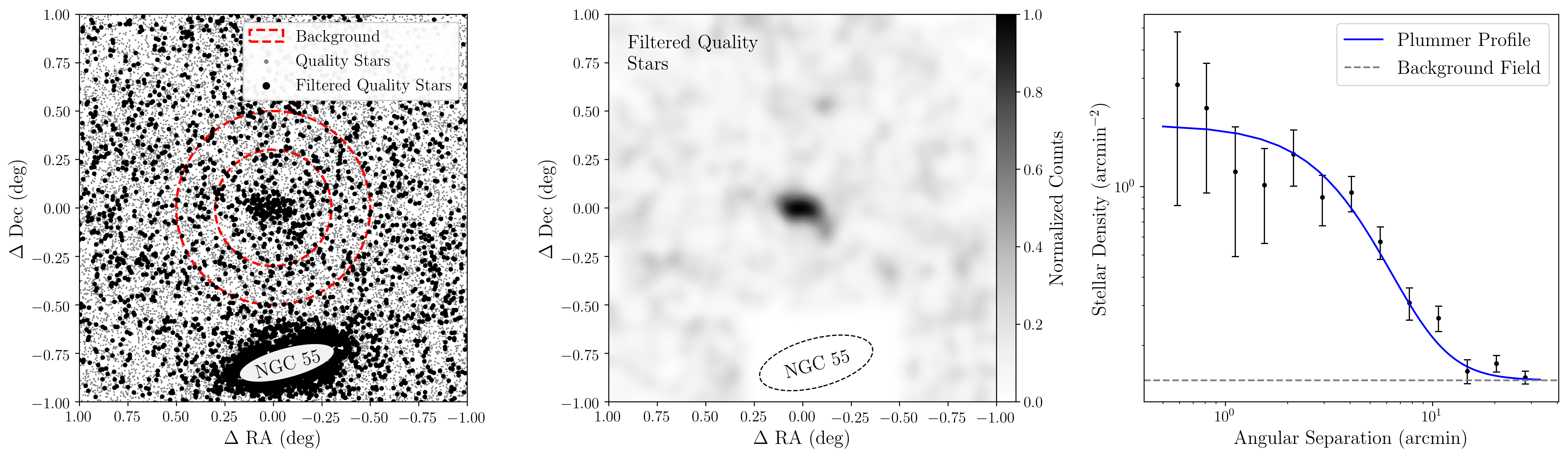}
\caption{Spatial distribution of stars near \name. 
\textit{Left}: Classified stars in our final dataset comprising DES, DELVE-DEEP, and one night of additional DECam follow-up (see \secref{cand_data}). 
Only stars which passed our quality cuts are shown, indicated with gray dots. 
Large black dots show stars additionally passing isochrone and color-color filters.
The red annulus indicates the region used to calculate the background stellar density. 
\textit{Middle}: A smoothed spatial map of stars passing quality cuts as well as passing isochrone and color-color filters. A small region containing NGC~55 has been removed for clarity; the approximate location of NGC~55 is indicated by the ellipse.
Note that for both the left and middle panels, the deepest data are only available in a $3 \deg^2$ region centered on \name, corresponding to the DECam field of view.
\textit{Right}: Radial density profile of stars passing all selection criteria, centered at the location of \name. 
The blue curve shows a Plummer profile with a scale radius of $a_h = 5.2\arcmin$ (\secref{cand_results}).
}
\label{fig:spatial}
\end{figure*}

The \simple search algorithm returned the locations of several thousand stellar density peaks. 
The distribution of these ``hotspots" falls steeply with increasing significance. 
In previous searches using the \simple algorithm for the detection of Milky Way satellites, most high-significance peaks coincide with real stellar systems, regions of small-scale spatial variance in extinction or stellar density, or survey artifacts \citep{Drlica-Wagner:2020}. 
Since this search is instead designed to identify distant, faint objects at the edge of detectability, the ratio of false positives is likely to be higher than in those searches. 
Therefore to filter through the large number of potential galaxy candidates, we applied extra layers of selection criteria for the hotspots. 
These selections, detailed in the following paragraphs, are designed to be relatively permissive, while still removing hotspots most likely to be spurious due to low statistical significance or their location in problematic areas of the survey footprint. 
A visual inspection of the remaining hotspots is necessary as a final step to determine any potential new candidate galaxies. 

First, we apply a significance threshold of $\sigma \geq 6.0$, consistent with previous applications of the \simple algorithm. 
Second, we apply a foreground mask encompassing several geometric criteria. 
Although we include the effects of reddening in the search algorithm, regions of high reddening tend to trace regions of high MW stellar density. In these regions the stellar field density and reddening often vary over small spatial scales, which can pose problems for the search algorithm in identifying real stellar overdensities. 
Therefore we restrict our search to regions of low interstellar extinction, $E(B-V) < 0.2$ \citep{Schlegel:1998}. Since the DES footprint primarily covers high-galactic latitude sky, this removes a negligible amount of sky area, $\roughly 0.2\deg^2$. 

We also masked regions near known astronomical objects that can cause spurious hotspots. 
These include nearby galaxies that are resolved into individual stars \citep{Corwin:2004,Nilson:1973,Webbink:1985,Kharchenko:2013,Bica:2008}, Milky Way globular clusters \citep[2010 edition]{Harris:1996}, open clusters (WEBDA)\footnote{\url{https://webda.physics.muni.cz}}, and bright stars \citep{YaleBSC}. 
We additionally masked regions around nearby MW satellite galaxies and other Local Group galaxies \citep{Drlica-Wagner:2020,McConnachie:2012} as well as overdensities in two narrow stellar streams, ATLAS \citep{Koposov:2014,Shipp:2018} and Phoenix \citep{Balbinot:2016, Tavangar:2022}.
For extended objects, the masked region covers the half-light radii of those objects, with a minimum masked radius of $0.05 \deg$. For bright stars and other objects without size information, a circular region of $0.1 \deg$ radius is masked. 
These masks cover a total of $96 \deg^2$, $\roughly 2.5\%$ of the DES footprint. 

\begin{figure*}
    \includegraphics[width=1.\textwidth]{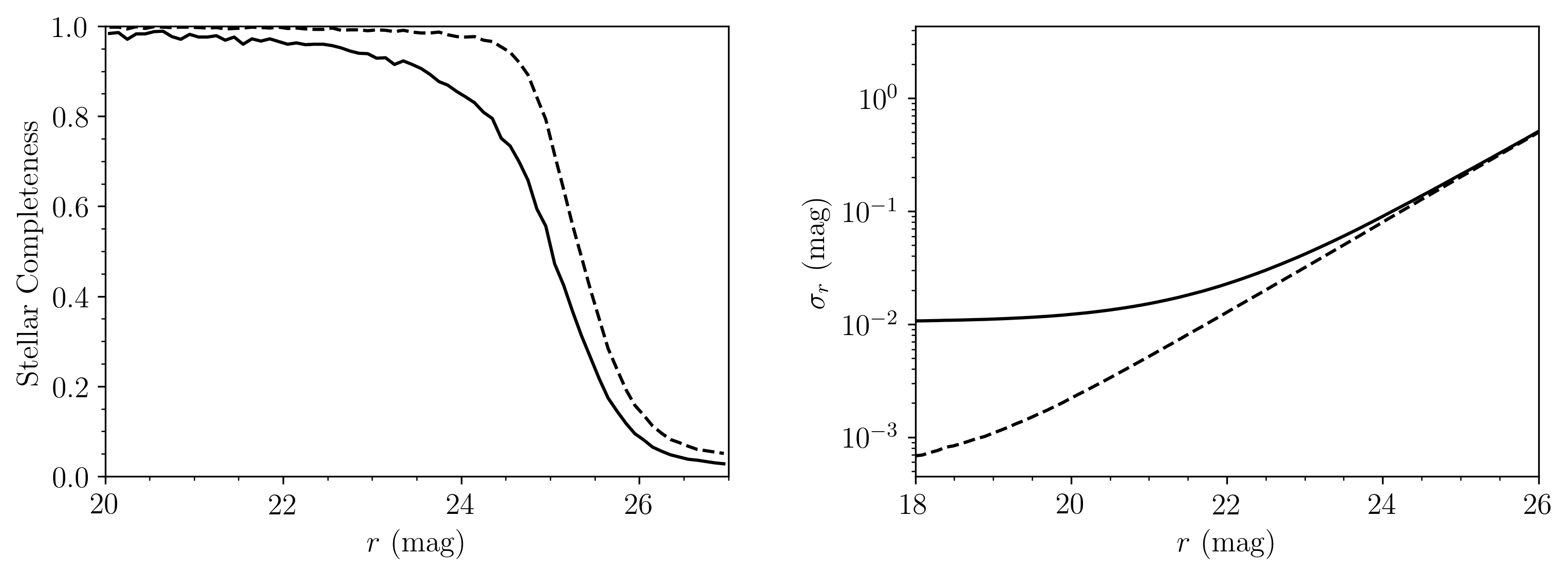}
    \caption{
    \textit{Left:} Stellar completeness model for the DES Y6 data set, validated against HSC-SSP DR1. 
    The dashed line indicates detection efficiency only, while the solid line indicates detection and correct star-galaxy classification. 
    \textit{Right:} Photometric uncertainty model for the DES Y6 data set. 
    The dashed line corresponds to statistical uncertainty alone, while the solid line includes a minimum photometric uncertainty of $0.01 \magn$.}
    \label{fig:completeness}
\end{figure*}

Applying these selections reduces the set of potential hotspots to $\roughly 300$. 
While this number far exceeds any expected number of true detections (see \secref{discussion_elvis}), a visual inspection of the results shows that many of these were obvious artifacts, predominantly due to data quality at the edge of the survey footprint. 
While the search algorithm does take into account survey coverage in the calculation of local stellar field density, the nature of the data quality at the outer border or near holes in the DES footprint led to many purportedly high-significance detections in these regions. 
These artifacts are easily identified and discarded.

After a visual inspection of diagnostic plots of spatial, color-magnitude, color-color, and morphological size distributions generated for each hotspot, we identified six previously undetected stellar systems which pass our selection criteria, only one of which stood out as highly significant and of particular interest. 
This system, \name, is the most statistically significant candidate identified in the search at $\sigma = 9.6$, exceeding our nominal threshold by several $\sigma$. 
Its large, diffuse nature makes it an extreme outlier in surface brightness and clearly distinguishes it upon visual inspection from common spurious overdensities picked up by the search algorithm which tend to be extremely compact. 
This unique structural profile, coupled with its proximity the the Local Volume galaxy NGC~55, make \name a particularly interesting target for further follow-up.
\figref{spatial} shows the spatial stellar distribution in the vicinity of \name, as well as an azimuthally-averaged radial profile centered on its location;
a more detailed analysis of the candidate galaxy is given \secref{cand}. 

The other systems had much lower Poisson significance, barely passing our nominal $\sigma \geq 6$ threshold. 
Furthermore, a search run over a previous version of the DES Y6 catalog identified a similar number of systems, but only two were found in both versions of the search.
One is the previously noted \name, which appeared at high significance in both searches.
The second system, located at $\left(\alpha, \delta\right) = \left(347.0, -2.0\right)$, sits very near the edge of the DES footprint and was identified at significance $\sigma = 6.02$.
Upon further investigation, the data quality was found to be inconsistent in this region of sky, making background estimation more difficult. 
Coupled with the fact that the remaining systems only appeared in one run of the search, we found it likely that these were spurious hotspots, and therefore chose to only consider \name for further investigation. 

\subsection{Sensitivity Analysis}
\label{sec:synthetic}

\begin{figure*}
    \includegraphics[width=1.\textwidth]{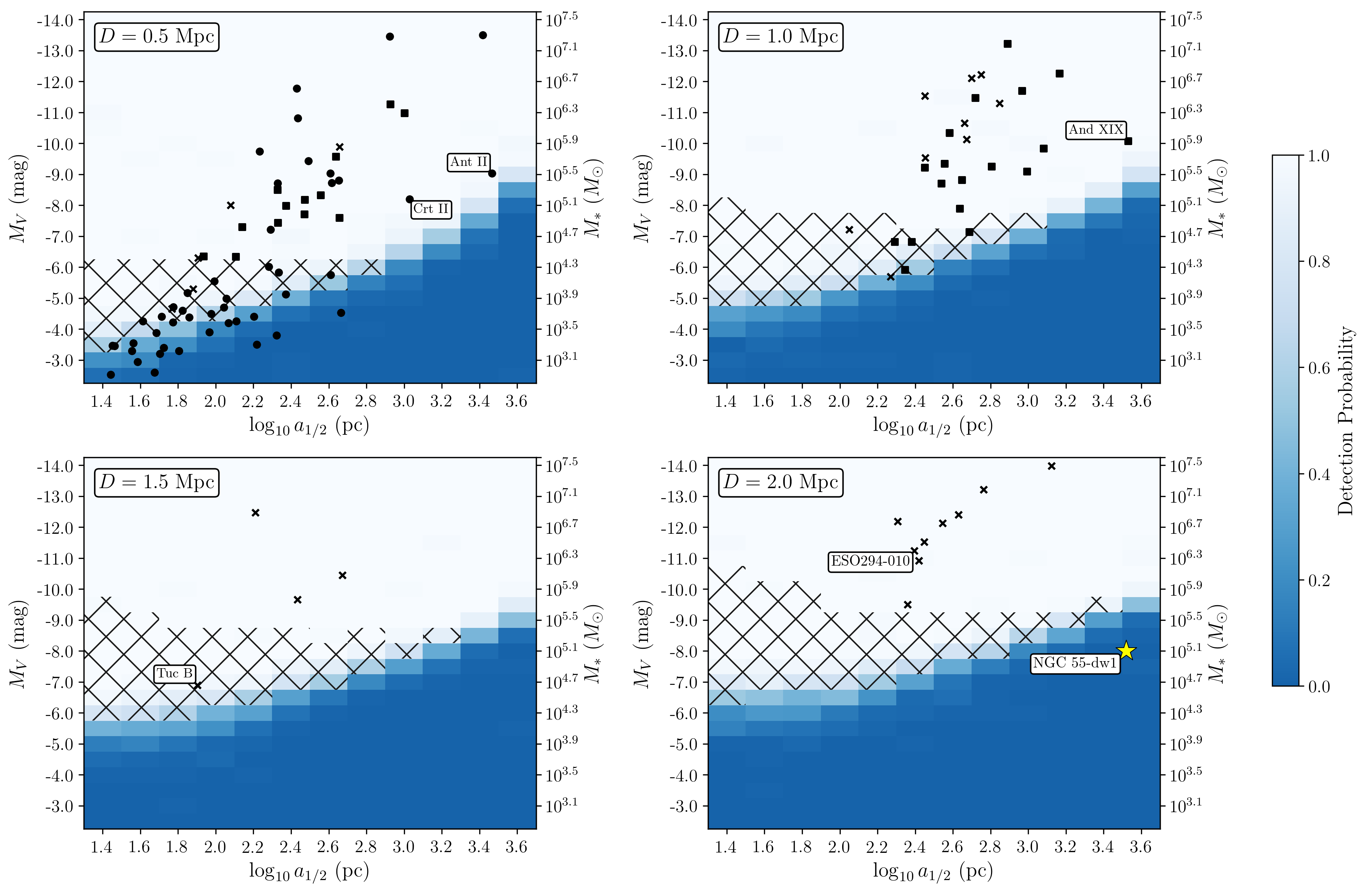}
    \caption{Expected detection efficiency of dwarf galaxies in our wide-area survey, evaluated by injecting simulated dwarf galaxies into the real DES data.
    The background color shows detection probability, defined as the fraction of simulated galaxies with Poisson significance $\sigma > 6$; 
    detection significance for galaxies brighter than $M_V = -12\magn$ is assumed without simulation. 
    The black hatched region contains galaxies which are likely undetectable due to blending effects in their central regions.
    \name is denoted with a large yellow star.
    Andromeda~XIX, Antlia~II, and Crater~II are the only known galaxies occupying a similar region of parameter space, and ESO~294-010 may be gravitationally associated with NGC~55 (\secref{discussion_ngc55}).
    Tucana B, located at $D=1.4\Mpc$ within the DES footprint, was not found by our search likely due to blending effects. 
    Sizes and luminosities of other Local Group galaxies are marked in the distance panel they are closest to: Milky Way satellites, all much nearer ($\lesssim 370$~kpc), with circles \citep{Drlica-Wagner:2020, Cerny:2021b, Cerny:2023}; M31 satellites with squares \citep{McConnachie:2012, Martin:2016}; and other Local Group galaxies with crosses \citep{McConnachie:2012, Sand:2022, Collins:2022, Collins:2023, McQuinn:2023, McQuinn:2023b}. 
    }
    \label{fig:sensitivity}
\end{figure*}

To estimate the sensitivity of our detection algorithm to distant dwarf galaxies in the DES Y6 data, we simulate galaxies with a range of luminosities and sizes at distances beyond the MW virial radius. These galaxies were simulated at the catalog level as collections of individually resolved stars. 
We assessed the stellar completeness of the DES Y6 catalog by comparison with data from Hyper Suprime Cam (HSC) SSP DR1 \citep{Aihara:2018}. 
In particular, we match the DES object catalog to stellar catalogs of the combined HSC Wide VVDS, Deep DEEP 2-3, and UltraDeep SXDS fields in a $\roughly 20 \deg^2$ region where the two datasets overlap. 
We calculate both the detection efficiency and stellar completeness as functions of DES $r$-band magnitude. The results are shown in the first panel of \figref{completeness}. 

We assigned photometric uncertainties on the simulated stellar magnitudes based on the depth and interstellar extinction at the location of each star according to the formula 
\begin{equation}
\label{eqn:photo_error}
\sigma_m = 0.01 + 10^{f(\Delta m)}.
\end{equation}
$\Delta m = m - m_{\rm lim}$ is the difference between the apparent magnitude $m$ of a star and the $10\sigma$ survey magnitude limit at the star's location $m_{\rm lim}$. The function $f(\Delta m)$ maps this difference to the median magnitude uncertainty. 
We derive $f$ by fitting the median magnitude uncertainty as a function of magnitude and magnitude limit. 
The photometric error model is shown in the second panel of \figref{completeness}. 
We impose a conservative 0.01\magn uncertainty minimum to ensure realistic representation of DES photometry for bright objects.

To generate realistic catalogs, we use probabilistic models for the spatial and flux distributions of the member stars of each synthetic galaxy. 
The spatial distribution of stars was sampled from a Plummer profile \citep{Plummer:1911}, and the initial masses were drawn from a \citet{Chabrier:2001} initial mass function (IMF). 
These have both been found to be good descriptions for known MW satellite galaxies \citep{Belokurov:2006, Sand:2010, Gennaro:2018a, Gennaro:2018b, Simon:2019}. 
A lower mass bound set to the hydrogen-burning limit of 0.08\Msun was imposed when sampling from the IMF. 
We used PARSEC isochrones \citep{Bressan:2012} to derive stellar photometry in the $g$, $r$, and $i$ bands from the initial stellar masses. We convert these absolute magnitudes into apparent magnitudes using the distance modulus of the simulated galaxy. 
Interstellar extinction was applied consistent with the real DES data as outlined in \secref{wide_data}.

We inject each galaxy's synthetic stellar catalog into the DES Y6 data and run the \simple search algorithm using the same search configuration used in the search over real data. 
In the typical use case of a search for undiscovered galaxies in real data, the algorithm scans over spatial location and distance modulus near an identified peak in stellar density to maximize its detection significance.
However, to save on computational time, we only search at the known spatial locations and distance moduli of the simulated galaxies. 
This ignores the possibility that background fluctuations or statistical fluctuations in the simulated stellar photometry could slightly increase the significance at different locations or distances, and therefore provides a conservative estimate of a galaxy's detectability. 
In a previous analysis of DES data using the \simple search algorithm, this choice only impacted the detection probability by at most a few percent for satellites close to the detection threshold \citep{Drlica-Wagner:2020}. 

The results of this analysis for galaxies located at heliocentric distances of 0.5, 1.0, 1.5, and 2.0\Mpc are shown in \figref{sensitivity}. At each distance slice, we simulate 100 galaxies for each $(a_{1/2}, \, M_V)$; \figref{sensitivity} represents $\approx 10^5$ total simulated systems. 
Note that galaxies with absolute magnitudes brighter than $M_V = -12 \magn$ were not fully simulated and instead assumed to be detected at the maximum significance. 

Due to its extreme size, the discovery of \name seems unexpected given our sensitivity at 2\Mpc. 
However, \name happens to lie in an area of sky where the local density of classified stars ($\roughly 1.2$ stars/arcmin$^{-2}$) is about 50\% lower than the mean across the footprint ($\roughly 1.8$ stars/arcmin$^{-2}$), resulting in \name being detected at high significance relative to the local background despite its diffuseness.
 
Since we inject galaxies at the catalog level, our predicted detection probabilities do not account for effects of blending in regions of high stellar density which may affect the detection and photometric measurement of member stars. 
For very bright or very compact galaxies, central stellar densities can exceed 100 stars/arcmin$^2$, roughly the density beyond which individual stars cannot be resolved by the DESDM pipeline \citep{Wang:2019}.
To assess the impact of blending on our sensitivity, we injected 175 galaxies at the image level, with absolute magnitudes $M_V \in [-8, -12]\magn$ and physical half-light radii $r_{1/2} \in [100, 1230]\pc$ at $D\roughly 2\Mpc$. 
We compute the discrepancy between each galaxy's absolute magnitude upon image-level injection and upon recovery by \code{SourceExtractor} as a function of density of injected stars brighter than $i = 27\magn$ within one half-light radius. 
The black hatches in \figref{sensitivity} denote regions of parameter space where this reduction in flux is likely to decrease the detection probability below 50\%. 
Galaxies in these regions were detected at high significance when injected at the catalog level; however, the likely non-detection of stars in their central regions due to blending would effectively dim the galaxies' magnitudes, decreasing their significance to below the detection threshold. 

Of note is the recently discovered Tucana B, an isolated ultra-faint dwarf galaxy at a distance $D=1.4 \Mpc$ \citep{Sand:2022}, which was not identified in our search despite being located within the DES footprint. 
Tucana B was discovered in a visual search of the DESI Legacy Imaging Surveys Data Release 9 \citep{Dey:2019}. 
As noted by the authors of the discovery paper, Tucana B is not well resolved in the DES $g$-band data and instead appears primarily as diffuse light. 
This is not particularly surprising, since as shown in \figref{sensitivity} Tucana B is compact enough relative to its magnitude that blending may cause difficulty for photometric measurements in its central region.
Since our search algorithm only identifies overdensities in the resolved stellar catalog, we are unable to find Tucana B and similarly unresolved objects. 

\section{Detailed analysis of \name}
\label{sec:cand}
 
The initial search over the DES data yielded one high significance candidate, \name.  
This candidate's distance, low surface brightness, and proximity to the Local Volume galaxy NGC~55 compelled us to acquire deeper imaging to confirm the detection and to obtain accurate measurements of distance, luminosity, and structural parameters. 
Here we describe the additional DECam imaging data used to characterize \name and the results of that analysis. 

\subsection{Deep Imaging for Candidate Characterization}
\label{sec:cand_data}

One piece of this deeper imaging comes from the DECam Local Volume Exploration (DELVE, \citealt{Drlica-Wagner:2021}), a wide-area survey that seeks to assemble contiguous DECam coverage of the southern sky.
One component of the DELVE survey, DELVE-DEEP, targets four isolated Magellanic Cloud analogue galaxies, including NGC~55.
DELVE-DEEP imaging covers the halos of each target to a $5\sigma$ depth of $g = 26.0\magn$ and $i = 25.0\magn$, and our galaxy candidate lies within the area imaged in this program as part of its observations of NGC~55.
This depth was achieved with $12 \times 300\second$ $g$-band exposures and $7 \times 300\second$ $i$-band exposures in addition to previous imaging done by DES. 
The DECam data taken as part of DELVE is processed consistently with the DES data through the DESDM pipeline. 

We also obtained dedicated DECam imaging for follow-up of \name; these observations were made during the first half of the nights of October 23 and 24, 2021 under Proposal ID 2021B-0307 (PI: Keith Bechtol). 
In order to improve our ability to characterize the system's distance, age, and metallicity, we observed in all of $gri$, taking $300\second$ exposures with three dither locations to alleviate the effects of chip gaps.
This set of observations added 25, 40, and 32 exposures in the $g$-, $r$-, and $i$- bands respectively. 
The exposures were processed through the same DESDM pipeline as the previous exposures taken as part of the DES and DELVE surveys. 
With the addition of the DELVE-DEEP observations near NGC~55 and our dedicated follow-up, we achieved a total exposure time of $\sim 3.5$ hours in each observing band, reaching $5\sigma$ depths of 26.8, 26.4, and 25.9\magn in each of $gri$ respectively.   

With the addition of these deeper data, it was necessary to have a stellar classifier which could be applied to the entire data set comprising all of DES, DELVE-DEEP, and further DECam follow-up.
The DELVE-DEEP catalog does not include a built-in stellar classifier; a stellar classifier has been developed for the wide DELVE DR2 public data release based on the \code{SourceExtractor} \code{SPREAD\_MODEL} quantity, but this classifier's efficiency falls steeply deeper than $g=22 \magn$ (see \citealt{Drlica-Wagner:2022}, Figure 8).
This classifier did not perform well when applied to the full deep data set, with the stellar sample dominated by galaxy contamination at the faint end. 
Therefore we created a custom morphological star-galaxy classifier to be used when fitting the candidate galaxy, in addition to the color-color criterion which was also used in the DES search.
We compared the distribution of the \code{BDF\_T} size parameter (see \secref{wide_data}) for all objects brighter than $i=26\magn$ in two fields, one with high stellar density very close to NGC~55 and one background field with relatively low stellar density. 
The first field, separated by $10\arcmin$ from NGC~55, overlapped the region of NGC~55 where its stars are individually resolved. The second field $60\arcmin$ from NGC~55 was distant enough that it was representative of the background in this region of sky. 
We derived the \code{BDF\_T} distribution of an approximately pure stellar sample by subtracting the \code{BDF\_T} distribution of the background from that of the high stellar density field. 
We found that a threshold of $\code{BDF\_T} < -0.02$ included $\roughly 80\%$ of the pure stellar sample, and we used this cutoff as a criterion to select a stellar sample for the entire region containing our candidate galaxy. 

\subsection{Analyzing deep follow-up imaging of \name}
\label{sec:cand_results}

\begin{figure}
    \centering
    \includegraphics[width=0.47\textwidth]{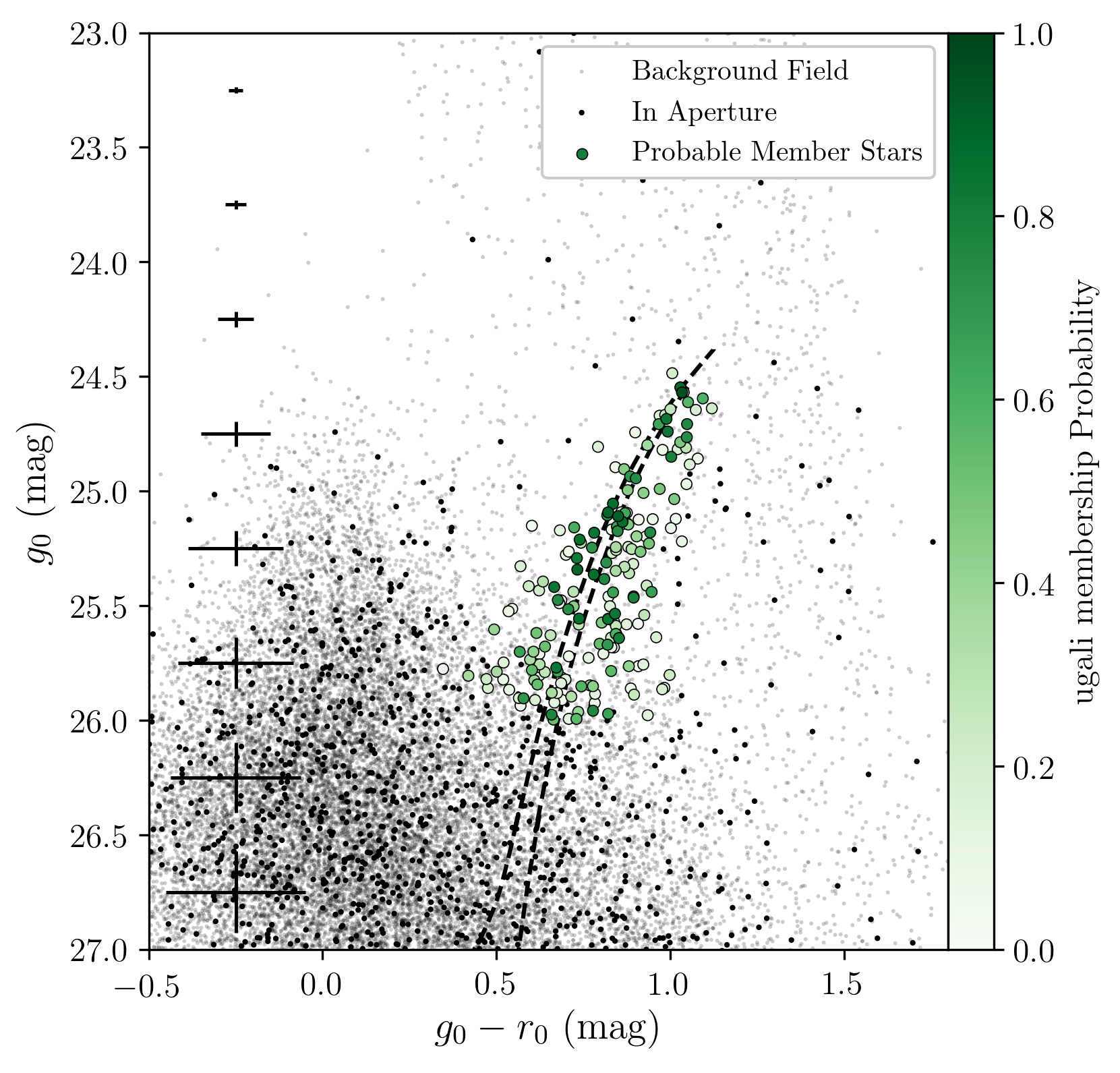}
    \caption{
    Reddening corrected color-magnitude diagram for the candidate dwarf \name identified in the search of DECam data. 
    All stars within a $0.5\deg$ radius ($\roughly 6 \, a_h$) are shown: field stars outside of the $6.6\arcmin$ best-fit spatial aperture in light gray, and stars within the aperture in black. 
    Stars brighter than a $g=26\magn$ limit with \code{ugali} membership probability greater than 5\% are colored by their respective probabilities. 
    The dashed curve shows a PARSEC isochrone \citep{Bressan:2012} for a stellar population at heliocentric distance $D=2.2\Mpc$ with age $\age=6.5 \Gyr$ and metallicity $\metal=0.00027$, the best-fit parameters from an MCMC fit -- 
    note that age and metallicity were not well constrained. 
    Median errors are indicated along the left side of the plot. 
    } 
    \label{fig:cmd}
\end{figure}

We re-ran the \simple algorithm over a small sky area covering the location of \name where we have additional data from DELVE-DEEP plus additional DECam follow-up. 
In the initial search, the candidate overdensity was most significant at the extreme distant end of the search space at a modulus of $26.5 \magn$; therefore, combined with having access to deeper data, we extended the allowed distance modulus parameter space an additional $1.0 \magn$. 
Shown in \figref{spatial} and \figref{cmd}, the result is a clearly visible overdensity with Poisson significance $\sigma=11.4$ relative to nearby background at $\mM=27.0\magn$, 
consistent with the tip of the red giant branch of an old, metal-poor population. 
Given that all but the brightest likely member stars are near the depth limit of our data, the somewhat broad distribution about the matching isochrone is consistent with a coherent stellar population accounting for our photometric errors. 

We obtained best-fit stellar population and morphological parameters for \name using the unbinned maximum likelihood formalism implemented in the Ultra-Faint Galaxy Likelihood (\ugali) software \citep{Bechtol:2015, Drlica-Wagner:2020}\footnote{\url{https://github.com/DarkEnergySurvey/ugali}}. 
We fit the system's color-magnitude distribution against a set of PARSEC \citep{Bressan:2012} isochrones assuming a Chabrier IMF \citep{Chabrier:2001}, and the 2D spatial stellar distribution was modelled with a Plummer profile. 
We simultaneously constrained the age (\age), metallicity (\metal), and distance modulus (\mM) of the isochrone, and the centroid location (\ra, \dec), semimajor axis (\major), ellipticity (\ellip), and position angle (\PA) of the Plummer profile by sampling the posterior distribution of each parameter using a Markov Chain Monte Carlo (MCMC) ensemble sampler (\emcee; \citealt{Foreman-Mackey:2013}). 

From these values we estimated the system's azimuthally averaged half-light radius (angular \rhalf and physical $r_{1/2}$), absolute V-band magnitude ($M_V$), average surface brightness within a half-light radius ($\mu$), total stellar mass ($M_*$), and mean metallicity (\feh). 
\tabref{properties} lists the properties of \name. 
This object's large physical size is especially noteworthy; \name is nearly the same size as the Small Magellanic Cloud (SMC). The only other faint galaxies with comparable sizes are Antlia~II and Andromeda~XIX, both $> 1 \magn$ brighter \citep{Torrealba:2019, Martin:2016}. 

\begin{deluxetable}{l c c }
\tabletypesize{\footnotesize}
\tablecaption{\label{tab:properties} Candidate Galaxy Properties}
\tablehead{\colhead{Parameter} & \colhead{Value} & \colhead{Units}}
\startdata
\ra & 3.874 $\substack{+0.015\\-0.014}$ & deg \\ 
\dec & -38.419 $\substack{+0.009\\-0.007}$ & deg \\ 
\mM & 26.71 $\substack{+0.05\\-0.12}$ & \magn \\ 
Distance & 2.20 $\substack{+0.05\\-0.12}$ & Mpc \\ 
\major & 5.2 $\substack{+1.2\\-0.8}$ & arcmin \\ 
\rhalf & 3.4 $\substack{+0.8\\-0.5}$ & arcmin \\ 
$r_{1/2}$ & 2.2 $\substack{+0.5\\-0.4}$ & kpc \\ 
\ellip & 0.56 $\substack{+0.10\\-0.12}$ & \ldots \\ 
P.A. & 156 $\substack{+7\\-8}$ & deg \\ 
$\tau$ & 6.5 $\substack{+4.3\\-2.7}$ & Gyr \\ 
$Z$ & 2.7 $(\substack{+2.8\\-0.6}) \times 10^{-4}$ & \ldots \\ 
\hline 
$M_V$ & -8.0 $\substack{+0.5\\-0.3}$ & \magn \\ 
$L_V$ & 2.4 $\substack{+0.8\\-1.0}$ & $10^5$ \Lsolar \\ 
$\mu$ & 32.3 & \magn arcsec$^{-2}$ \\ 
$M_*$ & 1.42 $\substack{+0.45\\-0.56}$ & $10^5$ \Msolar \\ 
{[Fe/H]} & -1.8 & dex \\ 
$l$ & 334.370 $\substack{+0.033\\-0.032}$ & deg \\ 
$b$ & -76.432 $\substack{+0.013\\-0.011}$ & deg \\ 
\enddata
\end{deluxetable}

\begin{figure}
    \centering
    \includegraphics[width=0.47\textwidth]{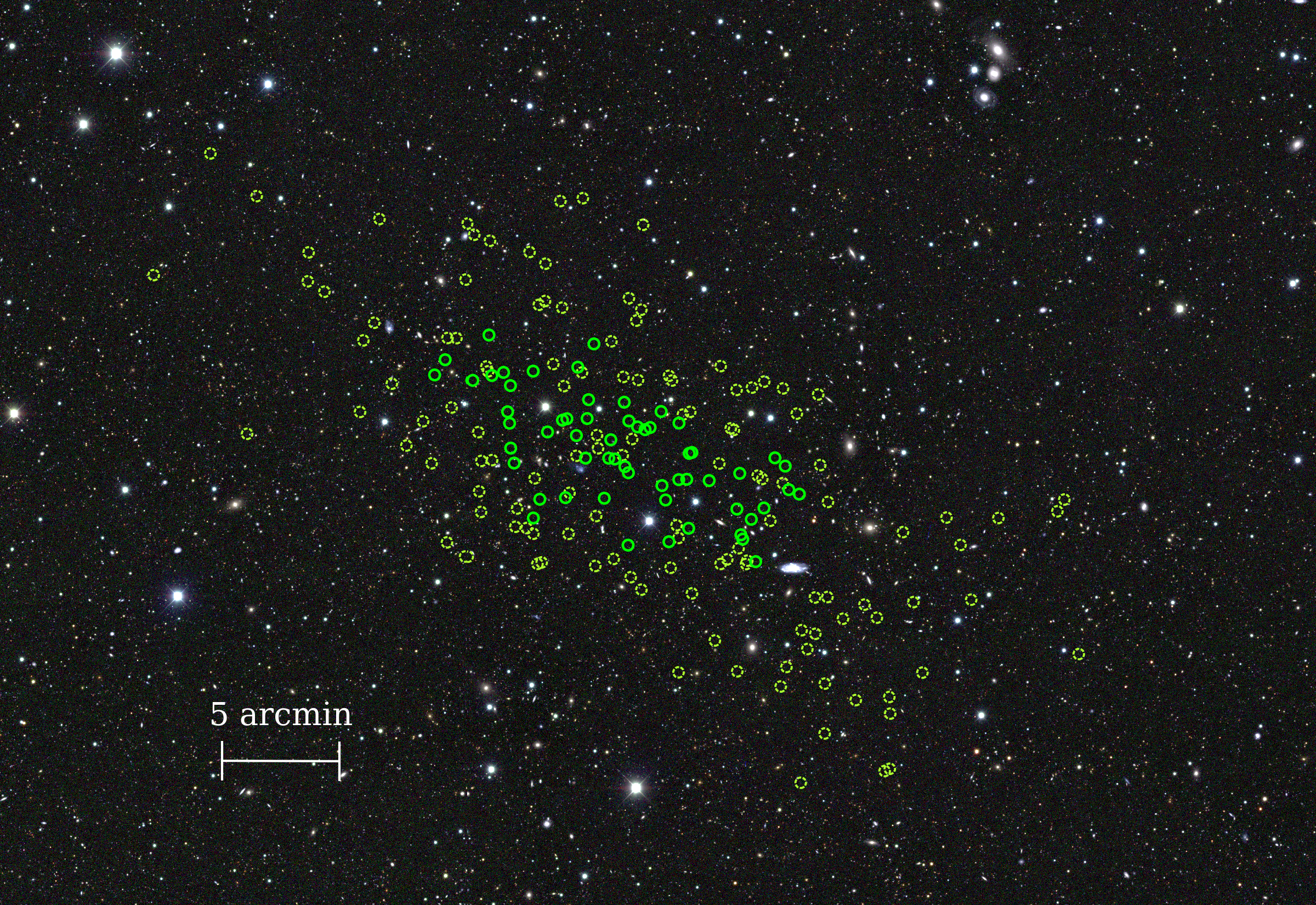}
    \caption{Coadd image of \name from our combined $gri$ DECam dataset with probable member stars indicated. 
    Higher confidence members with membership probability greater than 50\% are highlighted in bright green circles, while stars with membership probability greater than 5\% are shown in dashed circles.
    The member stars are all clearly resolved in our stellar catalog.}
    \label{fig:overlay}
\end{figure}

We also calculate the membership probability for each star based on its Poisson probability of belonging to \name based on its flux, photometric uncertainty, and spatial position, given an empirical model of the local background stellar population and of a candidate dwarf galaxy. 
Likely member stars above the magnitude limit used for our characterization analysis ($g=26\magn$) are shown overlayed on the sky image in \figref{overlay}. 
\appref{mcmc} shows the posterior probability distributions for each parameter in \figref{posteriors}.

Due to the relatively large photometric uncertainties for all but the brightest likely member stars, the age and metallicity of the stellar population are not well constrained by our fit. 
In comparison to other satellites in the Local Volume, our best-fit age of $6.5\substack{+4.3\\-2.7}\Gyr$ is very young. 
However, the age posterior is especially broad, including ages as old as $10.8\Gyr$ within the 68\% confidence interval and remaining roughly constant from $10.8\Gyr$ to $13.5\Gyr$. 
The metallicity distribution also skews heavily toward the metal-poor end of the parameter space, and \name therefore remains consistent with an old metal-poor stellar population.  
The distance modulus posterior displays a slight bimodality: the distribution is sharply peaked around $\mM = 26.7$ ($D=2.2\Mpc$), but with a small secondary hump near $\mM = 26.3$ ($D=1.8\Mpc$). 

\section{Discussion}
\label{sec:discussion}

\subsection{Interpretation of \name as a possible satellite galaxy of NGC~55}
\label{sec:discussion_ngc55}

Of particular note is our candidate galaxy's proximity to NGC~55, an LMC analogue approximately co-distant with \name.
The distance modulus of NGC~55 is measured to be 26.58\magn using the tip of the RGB method \citep{Tanaka:2011}; recent studies of Cepheid variables \citep{Gieren:2008} and blue supergiants \citep{Kudritzki:2016} lead to estimates of 26.43 and 26.85\magn respectively. 
Our candidate's best-fit distance modulus of $26.71\substack{+0.05\\-0.12}\magn$ sits comfortably within these estimates. 
The two galaxies are separated on the sky by approximately $47'$.
Assuming the two galaxies are both located at a heliocentric distance of 2.2\Mpc, they would be separated by a physical distance of only 30\kpc.  

The large spatial extent $r_{1/2}=2.2\substack{+0.5\\-0.4} \kpc$ relative to its $M_V=-8.0\substack{+0.5\\-0.3} \magn$ luminosity makes \name unusual among the known population of Local Volume dwarf galaxies. 
\name falls in a similar portion of parameter space to the MW satellites Antlia~II \citep{Torrealba:2019} and Crater~II \citep{Torrealba:2016a}, as well as the M31 satellites And~XIX, And~XXI, and And~XXIII \citep{Martin:2016}. 
Intense tidal stripping has been proposed as a formation mechanism for such large, diffuse galaxies \citep{Sanders:2018, Torrealba:2019, Collins:2019, Collins:2021}, and tidal features have been observed in Antlia~II and tentatively in Crater~II \citep{Ji:2021, Vivas:2022}.
Scl-MM-Dw2, a satellite of the Local Volume galaxy NGC~253, is brighter than \name ($M_V=-12\magn$) but has similar structural properties ($r_{1/2} = 2.9\kpc$, $\ellip=0.66$) and is being tidally disrupted by its host \citep{Toloba:2016, Mutlu-Pakdil:2022}. 
Our candidate's large, diffuse nature, high ellipticity, and proximity to its potential host galaxy are similarly suggestive of tidal interactions with NGC~55. 
While the small bimodality in the distance modulus posterior is likely attributable to photometric uncertainty, it could be interpreted as weak evidence of the presence of tidal features. 
However, recent $N$-body simulations have found that tidal effects alone may not be sufficient to explain the extreme sizes of these galaxies, indicating that they may not be bound systems at equilibrium or that their inner density profiles deviate significantly from expectations \citep{Errani:2022,Borukhovetskaya:2022}.
Due to the depth limitations of our ground-based optical imaging, confirmation of tidal stripping or measurement of stellar kinematics will require further follow-up. 

\begin{figure*}
    \centering
    \includegraphics[width=1.0\textwidth]{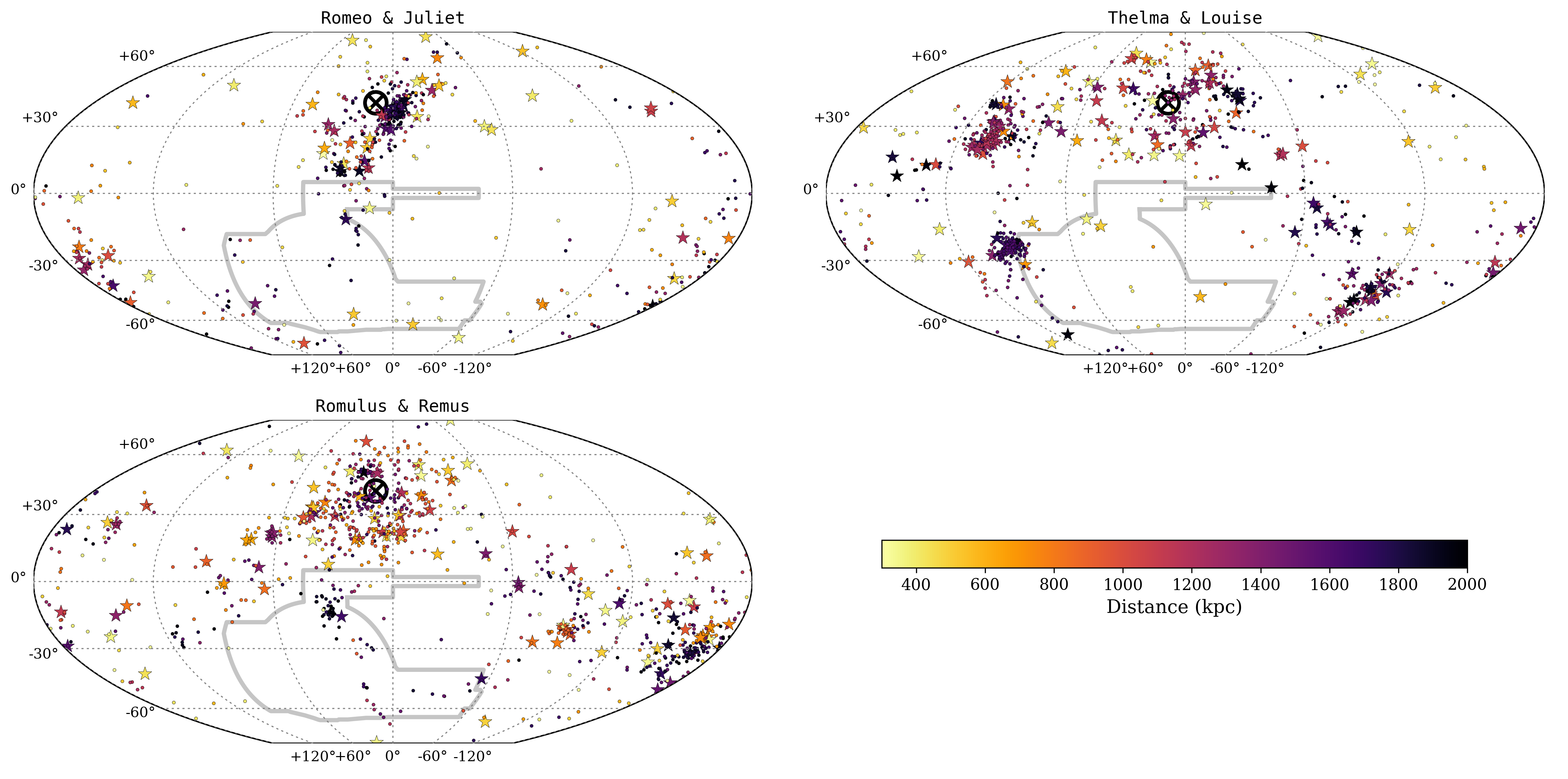}
    \caption{One realization of the nearby dwarf galaxy population for each of the three simulations from the ELVIS suite. 
    Each marker represents a galaxy predicted by applying the \cite{Nadler:2018iux} galaxy--halo connection model, with best-fit parameters derived from the MW satellite population \citep{Nadler:2020}, to the simulations' subhalo catalogs. 
    Galaxies are colored by heliocentric distance $D$, with larger stars indicating galaxies which should be detectable by DES if they were located in the footprint (gray outline). 
    Galaxies within the approximate virial radius of the M31 analogue ($300 \kpc$, large $\otimes$) are not shown;
    an arbitrary rotation about an axis through the M31 analogue's location could be made without changing its relative location to the DES footprint. 
    \textit{Top Left}: The relatively homogeneous \code{Romeo \& Juliet} simulation, with the line-of-sight chosen such that five detectable galaxies fall within the footprint. 
    Rotating about the M31 pole often leads to no detectable galaxies for this simulation. 
    \textit{Top Right}: \code{Thelma \& Louise} contains several spatially clustered groups of galaxies. 
    In this realization, one such cluster lies on the western edge of the DES footprint. Rotating this and other clusters into or out of the footprint has dramatic effects on the number of expected detectable galaxies. 
    \textit{Bottom Left}: In this realization of \code{Romulus \& Remus}, only one detectable galaxy falls inside the DES footprint.}
    \label{fig:elvis}
\end{figure*}

If \name is interpreted as a satellite of NGC~55 its discovery is not entirely surprising. 
By applying the best-fit galaxy--halo connection model from \citet{Nadler:2020} (described in \secref{discussion_elvis})  
to the LMC-mass zoom-in simulations from the Symphony compilation \citep{Nadler:2022dvo}, we predict that isolated LMC-mass halos host $\approx 1$ satellite of \name's absolute magnitude, on average, among a total satellite population of $\approx 10$ with $M_V \lesssim -3$, consistent with predictions from other studies (\eg, \citealt{Dooley:2017b}).
Thus, \name may naturally be interpreted as one of the brightest satellites of NGC~55, perhaps along with the $M_V \approx -11\magn$ ESO 294-010, separated from NGC~55 by 120\kpc \citep{Karachentsev:2002}. 
This scenario can be confirmed by precise distance measurements and a deep search for other galaxies in NGC~55's immediate vicinity.

\subsection{Wide-area search implications for total population of galaxies within 2~Mpc}
\label{sec:discussion_elvis}

To characterize the expected population of observable galaxies beyond the MW virial radius, we examine a set of high-resolution cosmological dark matter-only (DMO) zoom-in simulations of LG-like volumes. 
These simulations are a subset of the Exploring the Local Volume in Simulations (ELVIS) DMO suite, which contains pairs of MW--M31-size dark matter halos within a high-resolution volume spanning 2 to 5\Mpc with subhalos resolved down to peak virial masses $M_{\rm peak} = 6\times10^7 \Msun$ \citep{Garrison-Kimmel:2014}. 
For each of the three MW--M31 analogue systems we analyze (nicknamed \code{Romeo \& Juliet}, \code{Thelma \& Louise}, and \code{Romulus \& Remus}), we populate the $z=0$ dark matter halo and subhalo populations in each simulation with galaxies according the galaxy--halo connection model described in \citet{Nadler:2018iux,Nadler:2020}. 

This model flexibly extrapolates abundance-matching relations between galaxy luminosity, galaxy size, and halo properties (specifically, peak maximum circular velocity and virial radius at accretion) and accounts for the disruption of subhalos due to MW-mass central galaxies using the random forest classifier developed in \citet{Nadler:2018}. 
We evaluate the model using the best-fit parameters derived from DES and Pan-STARRS1 MW satellite observations in \cite{Nadler:2020}.\footnote{For any halos that are isolated (\ie, are not within the virial radius of a larger halo) at $z=0$, we neglect disruption by the central galaxy that would occupy the MW and M31 analogues; this choice has a negligible impact on the expected number of galaxies detectable in our DES search because the contribution of splashback halos to the total field population is small.}
Note that this galaxy--halo connection model is defined using halo properties defined at or before infall. Thus, although its parameters are fit to MW satellite data, we expect the model to apply to the population of field dwarfs surrounding the MW; moreover, the underlying abundance-matching model is calibrated on the GAMA field luminosity function \citep{Loveday150501003} at the bright end ($M_r<-13~\magn$; \citealt{Nadler:2018iux}).

We note the existence of more recent hydrodynamic simulations of LG-like systems, expanding on the ELVIS suite by applying Feedback In Realistic Environments (FIRE, \citealt{Hopkins:2014, Garrison-Kimmel:2017, Hopkins:2018, Garrison-Kimmel:2019a, Garrison-Kimmel:2019b}). 
However, our galaxy--halo connection model was calibrated on DMO simulations, and populating the DMO suite with synthetic galaxies via our model allows us to consider fainter systems than those resolved in the hydrodynamic versions of ELVIS, which have baryonic particle masses ranging from $3500$ to $7000~\Msun$ and thus only resolve galaxies down to stellar masses of $\approx 10^5~\Msun$. 
Sufficiently high-resolution hydrodynamic simulations of MW-like environments out to $\roughly 750 \kpc$ do exist \citep{Applebaum:2021}, and extensions of such simulations encompassing the entire LG are an important avenue for future works.

In order to characterize the detectability of the synthetic galaxies by DES, we first orient the population relative to the two most massive halos in the simulation, which we respectively map onto M31 and the MW.
We take the coordinates of the MW-analogue host as the observer location, and measure the distance to each (sub)halo relative to this origin. 
To correctly capture the influence of the location of the M31-analogue on the isolated halo populations, we place the DES footprint on the sky relative to the location of the M31-analogue host to mirror their relative locations in the actual Local Group.

For the galaxies located inside the DES footprint and within our nominal distance range of $300 \kpc - 2 \Mpc$, we produce mock stellar catalogs derived from the luminosity, physical size, and distance of each galaxy by the same procedure described in \secref{synthetic}.
We inject this mock stellar catalog into the real DES Y6 data at their appropriate location on the sky, and pass this data through the \simple search algorithm.
Note that due to computational limits we search only at the simulated galaxies' known locations and distances. 

\begin{deluxetable}{lcccc}
\tabletypesize{\footnotesize}
\tablecaption{\label{tab:elvis} Summary Statistics of Detectable Galaxies in ELVIS simulations given our DES Search Sensitivity}
\tablehead{\colhead{Simulation} & \colhead{Range} & \colhead{Median} & \colhead{Mean} & \colhead{\thead{Standard \\ Deviation}}}
\startdata
\code{Romeo \& Juliet} & $0-6$ & 3 & 2.6 & 1.6 \\
\code{Thelma \& Louise} & $0-42$ & 8 & 11.6 & 11.2 \\
\code{Romulus \& Remus} & $0-14$ & 5.5 & 6.2 & 3.3 \\
\enddata
{\footnotesize \tablecomments{For each simulation, we place the mock DES footprint in 60 different sky locations corresponding to 60 equally spaced rotations about the analogue MW--M31 axis. Shown here is the range, median, mean, and standard deviation about the mean of the number of detectable galaxies within the footprint among those realizations.}}
\end{deluxetable}

There is wide variation in the predicted detectable distant dwarf galaxy population among the three MW--M31 analogue systems and among particular realizations of any given system. For one system, \code{Romeo \& Juliet}, the distribution of field galaxies outside of the mock MW and M31 virial radii is relatively uniform; for the other two systems, the distribution is inhomogeneous, with galaxies often clustered together spatially, particularly in the case of \code{Thelma \& Louise}.  
Although we can fix the DES footprint correctly relative to the location of the M31-analogue, this still leaves an arbitrary rotation about the MW--M31 axis to determine the sky area covered by the footprint. 
Due to their spatial inhomogeneity, the number of galaxies inside the DES footprint is highly dependent on this chosen line of sight. 
A summary of the detectable galaxy population is given in \tabref{elvis}, and one chosen realization for each of the three simulations is demonstrated in \figref{elvis}. 

\begin{deluxetable}{lcccc}
\tabletypesize{\footnotesize}
\tablecaption{\label{tab:LG} Local Group galaxies between $D=300\kpc$ and $D=2 \Mpc$ inside the DES footprint}
\tablehead{\colhead{Name} & \colhead{Distance (kpc)}& \colhead{$M_V$ (mag)} & \colhead{$r_{1/2}$ (pc)} & \colhead{$\sigma$}}
\startdata
Eridanus II & 370 & -7.1 & 277 & 37.5 \\
Phoenix & 420 & -9.9 & 454 & 37.5 \\
IC 1613 & 760 & -15.2 & 1496 & 37.5 \\
Tucana & 890 & -9.5 & 284 & 23.4 \\
Tucana B & 1400 & -6.9 & 80 & - \\
ESO 410-005G & 1920 & -11.5 & 280 & 10.9 \\
\enddata
{\footnotesize \tablecomments{ $\sigma$ is the Poisson significance (maximum 37.5) recovered by the \simple algorithm during our wide-area search of DES Y6. Properties for Eridanus II are taken from \citet{Crnojevic:2016b} and \citet{Martinez-Vazquez:2021}; Tucana B from \citet{Sand:2022}; all other galaxies from an updated catalog of \citet{McConnachie:2012}.}}
\end{deluxetable}

Due to the broad range of detectable galaxies predicted by our realizations of the ELVIS simulations and the small number of systems that pass the detection thresholds, it is difficult to draw strong conclusions about the galaxy--halo connection based on our search. 
However, we note that our result of only one newly discovered galaxy likely located just beyond our nominal $300 \kpc - 2 \Mpc$ search range is statistically consistent with all three mock LG analogue simulations.  
The Local Group contains six known galaxies within this distance range visible in the DES footprint, listed in \tabref{LG}. 
The system with the most homogeneous population, \code{Romeo \& Juliet}, predicts the lowest number of detectable LG field dwarfs (median 3), most closely matching our search results. 
In the case of the most inhomogeneous system, \code{Thelma \& Louise}, some lines of sight led to a much higher expectation (as many as 42) for the number of detectable galaxies; however, the variation is so high that even a null result would have been consistent with the prediction. Reducing this uncertainty using constrained simulations, which are tailored to match the specific environment of the LG (\eg, \citealt{Carlesi:2016qqp,Libeskind:2020,Sawala:2022}), is therefore a valuable avenue for future work.

There is also significant theoretical uncertainty in the faint-end galaxy--halo connection itself (see \eg, \citealt{Bullock:2017xww,Wechsler:2018pic,Sales:2022ich} for reviews).
Here, we fixed our galaxy--halo connection model to the best fit derived using the MW satellite population in \cite{Nadler:2020}, and therefore neglected the associated theoretical uncertainty in the number of predicted detectable galaxies. 
It will be interesting to refine this analysis in order to test whether the galaxy--halo connection inferred from MW satellites is consistent with that inferred from isolated galaxies throughout the LG. 

\subsection{Outlook}

This search pushed the limits of current, wide, ground-based imaging sensitivity. We used the full six-year DES dataset with a search algorithm redesigned for use in the context of distant, compact dwarf galaxies, and still needed additional follow-up data and a specialized stellar classification metric to confirm the existence of a galaxy at the faintest end of our search. 
Pushing any further out in distance is generally limited by star-galaxy confusion and photometric uncertainties, and searching for undiscovered galaxies by identifying spatial stellar overdensities is further hampered by blending effects and variation in the apparent density of stars due to variation in the depth and delivered image quality across the footprint at the faint end. 

However, the next generation of telescopes coming online in the very near future will greatly expand our ability to obtain a full census of the ultra-faint dwarf galaxy population in the nearby universe. Two space-based observatories, the visible to near-infrared Euclid \citep{Racca:2016, Euclid:2022a} and the Nancy Grace Roman Space Telescope (formerly WFIRST, \citealt{Akeson:2019}), are set to begin observations within the next five years. 
The unprecedented combination of survey area, depth, spatial resolution, and low sky background of these surveys will be able to deliver precise astrometric measurements for faint sources \citep{WFIRST:2019}, improved detection of linear structure of resolved stars \citep{Pearson:2022}, improved detection of extended low surface brightness structure \citep{Euclid:2022b}, and competitive constraints on the Milky Way's dark matter subhalo population via its microlensing signatures \citep{Pardo:2021}.
These improvements will contribute significantly to our ability to discover and characterize the remaining dwarf galaxy population of the Local Group and to our understanding of the dark matter substructure which these galaxies inhabit. 

The ground-based Vera C. Rubin Observatory (Rubin) Legacy Survey of Space and Time (LSST) will follow directly in the footsteps of DES, imaging the entire southern sky in visible to near-infrared $ugrizy$ filters, expected to reach a 5-$\sigma$ point source image depth of ($g$, $r$, $i$) = (27.4, 27.5, 26.8) in the visible bands after the 10-year survey \citep{Ivezic:2019}. 
\citet{Mutlu-Pakdil:2021} presents a preview of the faint dwarf galaxy discoveries that will be possible in the next decade with the Vera C. Rubin Observatory and Subaru Hyper Suprime-Cam by combining images from the Panoramic Imaging Survey of Centaurus and Sculptor (PISCeS, \citealt{Sand:2014, Toloba:2016, Crnojevic:2016, Crnojevic:2019, Hughes:2021}) with extensive image simulation.
In particular, next generation deep surveys will be able to resolve horizontal branch stars of galaxies within a fiducial distance of 1.5\Mpc, uncovering low-surface brightness systems down to $\mu_{V, 0} \roughly 30$\magn arcsec$^{-2}$ for $M_V = -6$ and $\mu_{V, 0} \roughly 29$\magn arcsec$^{-2}$ for $M_V = -5$. 
Similar depth surveys will be able to reach $\roughly 2\magn$ below the tip of the red giant branch for systems as distant as $5\Mpc$, enabling a secure census of dwarf galaxies brighter than $M_V \approx -8 \magn$. 
Furthermore, a matched-filter search technique similar to the one employed in this paper remains powerful for identifying ultra-faint systems. 

In addition to undiscovered ultra-faint dwarfs, recent studies also predict an undetected population of isolated, spatially extended ultra-diffuse galaxies ($M_V \lesssim -8$, $\mu \gtrsim 24\magn\,\textrm{arcsec}^{-2}$) within the LG, which should also be found by a full-sky survey with the sensitivity of LSST \citep{Newton:2023}.
Studies of faint systems outside the MW and beyond the LG are necessary to build up a robust sample of galaxies populating dark matter halos from a broad range of environments. 
Detecting and accurately characterizing these systems is needed to fully test the astrophysics relevant to the formation and evolution of dwarf and diffuse galaxies (\eg, baryonic feedback, tidal stripping, reionization, \etc) and to test the $\Lambda$+Cold Dark Matter (\LCDM) model on increasingly smaller scales. 
There is likely a large population of such systems awaiting discovery, and near-future studies will continue to push the frontier of discovery deeper. 

\section{Summary}

We performed a search over the DES Y6 data for faint field dwarf galaxies with heliocentric distances $D = 0.3 - 2 \Mpc$ using the \simple matched-filter search algorithm. This algorithm identifies galaxies as arcminute-scale overdensities of individually resolved stars. 
We assessed the completeness of our search by the injection and recovery of synthetic galaxies inserted into the DES data at the catalog level, with a small number of galaxies being inserted at the image level to assess blending effects. 
For smaller ultra-faints (physical half-light radius $\lesssim 100\pc$), we expect completeness to roughly $M_V = -6.5\magn$ for galaxies with $D = 0.5\Mpc$ and $M_V = -10.5\magn$ for galaxies with $D = 2\Mpc$. 
For larger galaxies (physical half-light radius $\gtrsim 1000\pc$), we expect completeness to roughly $M_V = -8.5\magn$ for galaxies with $D \leq 1.0\Mpc$ and $M_V = -10.0\magn$ for galaxies with $D = 2\Mpc$. 

We do not find any new dwarf galaxies within our search space.
Based on a set of high-resolution cosmological zoom-in simulations of LG-like volumes, this result is not entirely inconsistent with expectations despite these simulations often predicting the existence of several detectable galaxies visible to our survey. 
With the exception of the unresolved Tucana B, we do recover the known galaxies within our search volume at high significance. 

We do detect a high-confidence galaxy just beyond our nominal search bounds. 
We report the discovery of \name, an ultra-diffuse galaxy located at $D=2.2 \Mpc$ with absolute magnitude $M_V=-8.0 \magn$ and azimuthally averaged physical half-light radius $r_{1/2} = 2.2 \kpc$. 
We obtained deep follow-up DECam imaging to confirm the system and measure its properties. 
This is the largest, most diffuse galaxy known at this luminosity. 
It is separated by only $47\arcmin$ from the LMC-mass NGC~55; assuming the two are roughly co-distant, they are separated by only 30\kpc. 
\name's proximity to a more massive host may explain its extreme structural properties; tidal interactions are a possible explanation for its large size, high ellipticity (\ellip = 0.56), and extremely low surface brightness ($\mu=32.3 \magn$ arcsec$^{-2}$). 

Future wide-area surveys such as the Rubin Observatory LSST will continue to fill the gaps in our knowledge of the ultra-faint dwarf population of the Local Volume. 
The continued discovery and study of these galaxies in nearby and distant environments will play an important role in our understanding of the nature of dark matter and of the assembly history of our local corner of the Universe. 

\section*{Acknowledgements}
MM and KB acknowledge support from NSF grant AST-2009441. 
Research by DC is supported by NSF grant AST-1814208.
ABP is supported by NSF grant AST-1813881.
JLC acknowledges support from NSF grant AST-1816196.
CEMV is supported by the international Gemini Observatory, a program of NSF’s NOIRLab, which is managed by the Association of Universities for Research in Astronomy (AURA) under a cooperative agreement with the National Science Foundation, on behalf of the Gemini partnership of Argentina, Brazil, Canada, Chile, the Republic of Korea, and the United States of America.
DJS acknowledges support from NSF grant AST-1821967 and AST-2205863.
JAC-B acknowledges support from FONDECYT Regular N 1220083. 

This project used data obtained with the Dark Energy Camera (DECam), which was constructed by the Dark Energy Survey (DES) collaboration.
Funding for the DES Projects has been provided by the U.S. Department of Energy, the U.S. National Science Foundation, the Ministry of Science and Education of Spain, 
the Science and Technology Facilities Council of the United Kingdom, the Higher Education Funding Council for England, the National Center for Supercomputing 
Applications at the University of Illinois at Urbana-Champaign, the Kavli Institute of Cosmological Physics at the University of Chicago, 
the Center for Cosmology and Astro-Particle Physics at the Ohio State University,
the Mitchell Institute for Fundamental Physics and Astronomy at Texas A\&M University, Financiadora de Estudos e Projetos, 
Funda{\c c}{\~a}o Carlos Chagas Filho de Amparo {\`a} Pesquisa do Estado do Rio de Janeiro, Conselho Nacional de Desenvolvimento Cient{\'i}fico e Tecnol{\'o}gico and 
the Minist{\'e}rio da Ci{\^e}ncia, Tecnologia e Inova{\c c}{\~a}o, the Deutsche Forschungsgemeinschaft and the Collaborating Institutions in the Dark Energy Survey. 

The Collaborating Institutions are Argonne National Laboratory, the University of California at Santa Cruz, the University of Cambridge, Centro de Investigaciones Energ{\'e}ticas, 
Medioambientales y Tecnol{\'o}gicas-Madrid, the University of Chicago, University College London, the DES-Brazil Consortium, the University of Edinburgh, 
the Eidgen{\"o}ssische Technische Hochschule (ETH) Z{\"u}rich, 
Fermi National Accelerator Laboratory, the University of Illinois at Urbana-Champaign, the Institut de Ci{\`e}ncies de l'Espai (IEEC/CSIC), 
the Institut de F{\'i}sica d'Altes Energies, Lawrence Berkeley National Laboratory, the Ludwig-Maximilians Universit{\"a}t M{\"u}nchen and the associated Excellence Cluster Universe, 
the University of Michigan, the NSF’s National Optical-Infrared Astronomy Laboratory, the University of Nottingham, The Ohio State University, the University of Pennsylvania, the University of Portsmouth, 
SLAC National Accelerator Laboratory, Stanford University, the University of Sussex, Texas A\&M University, and the OzDES Membership Consortium.

The DELVE project is partially supported by Fermilab LDRD project L2019-011 and the NASA Fermi Guest Investigator Program Cycle 9 No. 91201. 

Based on observations at Cerro Tololo Inter-American Observatory, NSF’s National Optical-Infrared Astronomy Laboratory (2012B-0001, PI: J. Frieman; 2019A-0305, PI: A. Drlica-Wagner; 2021B-0307, PI: K. Bechtol) which is operated by the Association of 
Universities for Research in Astronomy (AURA) under a cooperative agreement with the National Science Foundation.

The DES data management system is supported by the National Science Foundation under Grant Numbers AST-1138766 and AST-1536171.
The DES participants from Spanish institutions are partially supported by MINECO under grants AYA2015-71825, ESP2015-66861, FPA2015-68048, SEV-2016-0588, SEV-2016-0597, and MDM-2015-0509, 
some of which include ERDF funds from the European Union. IFAE is partially funded by the CERCA program of the Generalitat de Catalunya.
Research leading to these results has received funding from the European Research
Council under the European Union's Seventh Framework Program (FP7/2007-2013) including ERC grant agreements 240672, 291329, and 306478.
We  acknowledge support from the Brazilian Instituto Nacional de Ci\^encia
e Tecnologia (INCT) e-Universe (CNPq grant 465376/2014-2).

We use simulations from the FIRE-2 public data release \citep{Wetzel:2023}. The FIRE-2 cosmological zoom-in simulations of galaxy formation are part of the Feedback In Realistic Environments (FIRE) project, generated using the Gizmo code \citep{Hopkins:2015} and the FIRE-2 physics model \citep{Hopkins:2018}.

This work used data from the Symphony suite of simulations (\href{http://web.stanford.edu/group/gfc/symphony/}{http://web.stanford.edu/group/gfc/symphony/}).

This manuscript has been authored by Fermi Research Alliance, LLC under Contract No. DE-AC02-07CH11359 with the U.S. Department of Energy, Office of Science, Office of High Energy Physics. 

\textit{Facility:} Blanco

\textit{Software:} \code{Astropy} \citep{Astropy:2013}, \code{emcee} \citep{Foreman-Mackey:2013}, \code{HEALPix} \citep{Gorski:2005}\footnote{\url{https://healpix.sourceforge.ne}}, \code{healpy}\footnote{\url{ https://github.com/healpy/healpy}}, \code{matplotlib} \citep{Hunter:2007}, \code{numpy} \citep{vanderWalt:2011}, \code{Scikit-Learn} \citep{Pedregosa:2011}, \code{scipy} \citep{Jones:2001}, \code{ugali} \citep{Bechtol:2015}\footnote{\url{https://github.com/DarkEnergySurvey/ugali}}.

\bibliographystyle{apj}
\bibliography{main}

\appendix

\section{Posterior Distributions of \name Parameters}
\label{app:mcmc}

\begin{figure*}[h!]
    \centering
    \includegraphics[width=1.0\textwidth]{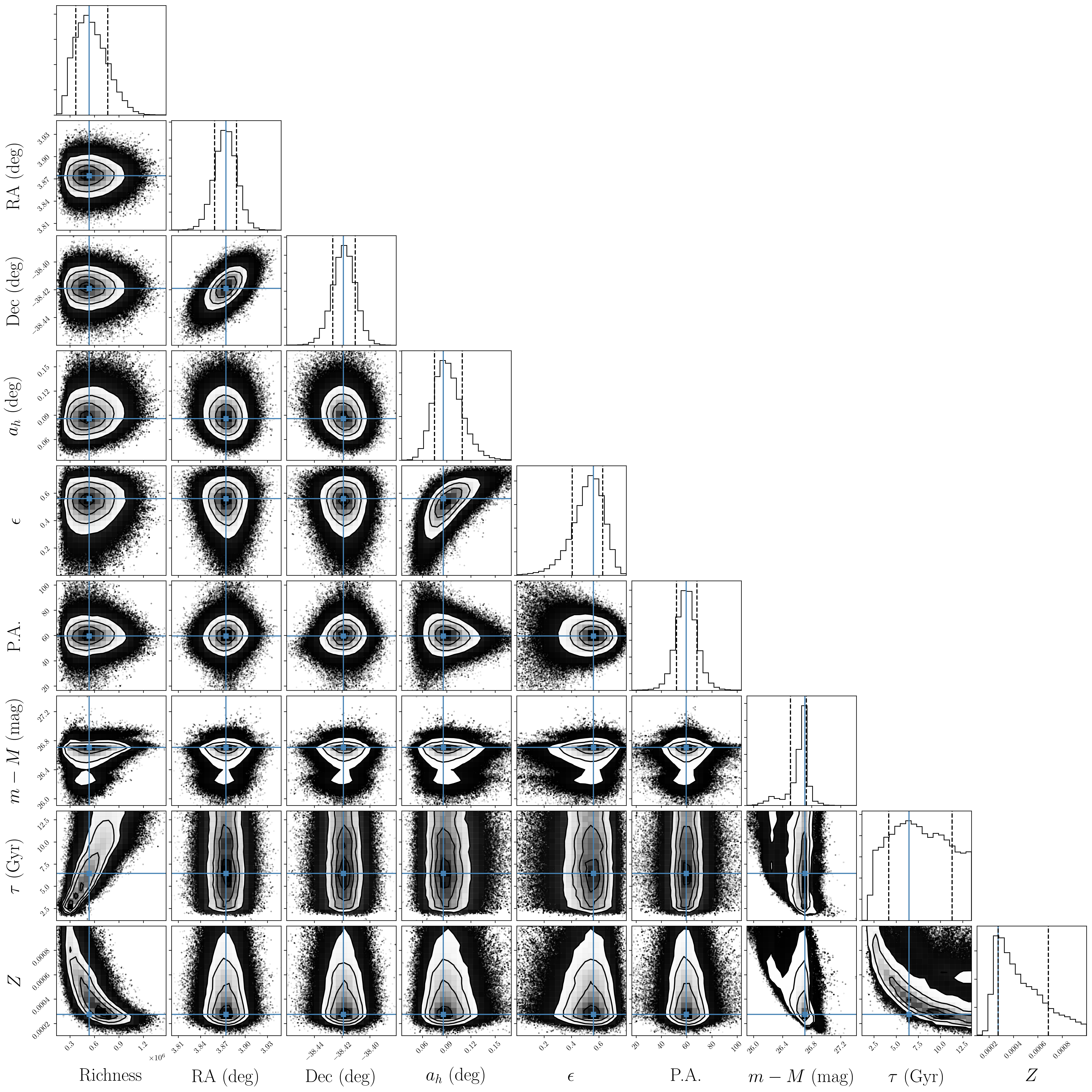}
    \caption{Posterior probability distributions for each parameter fit by the MCMC sampling described in \secref{cand_results}. Age and metallicity are not well constrained due to lack of precise photometry for stars beyond the magnitude limit of our dataset.}
    \label{fig:posteriors}
\end{figure*}

\end{document}